\documentclass[%
 aip,
 floatfix,
 amsmath,amssymb,
 reprint,%
]{revtex4-1}

\usepackage[mathcal]{eucal}     %for curly F's for Fourier Transform.
\usepackage{graphicx}% Include figure files
\usepackage{dcolumn}% Align table columns on decimal point
\usepackage{bm}% bold math
\usepackage{epigraph} %for twee quotes
\usepackage[normalem]{ulem}

\usepackage[utf8]{inputenc}
\usepackage[T1]{fontenc}
\usepackage{mathptmx}
\usepackage{etoolbox}

\usepackage[T1]{tipa}
\usepackage[tipa]{ucs}
\usepackage[dvipsnames]{xcolor}
\definecolor{greige}{RGB}{233, 216, 166}
\definecolor{hotpink}{RGB}{172, 41, 123}
\definecolor{RAFblue}{RGB}{94, 125, 158}
\definecolor{mynavy}{RGB}{16, 60, 111}

\usepackage{tikz}
\usetikzlibrary{shapes, arrows.meta, positioning}
\tikzstyle{startstop} = [rectangle, rounded corners, minimum width=3cm, minimum height=1cm,text centered, text width=5cm, draw=black, fill=hotpink!60] 
\tikzstyle{startstop2} = [rectangle, rounded corners, minimum width=3cm, minimum height=1cm,text centered, text width=3cm, draw=black, fill=hotpink!60]
\tikzstyle{startstopPALE} = [rectangle, rounded corners, minimum width=3cm, minimum height=1cm,text centered, text width=4cm, draw=black, fill=hotpink!40]
\tikzstyle{startstop3} = [rectangle, rounded corners, minimum width=3cm, minimum height=1cm,text centered, draw=black, fill=hotpink!60]
\tikzstyle{GSbox} = [rectangle, rounded corners, minimum width=0.5cm, minimum height=1cm,text centered, text width=3cm, draw=black, fill=hotpink!20] 
\tikzstyle{io} = [rounded rectangle, minimum width=3cm, minimum height=1cm, text centered, text width=5cm, draw=black, fill=RAFblue!40]
\tikzstyle{decision} = [diamond, minimum width=3cm, minimum height=1cm, text centered, text width=2cm, draw=black, fill=RAFblue!70]
\tikzstyle{arrow} = [thick,->,>=stealth]

\usepackage{multirow}

\usepackage[normalem]{ulem} %for strikeout text  \sout
\usepackage[colorlinks]{hyperref}
\AtBeginDocument{\hypersetup{allcolors=[rgb]{0.41,0.59,0.55}}}

\usepackage{caption}
\usepackage{subcaption}

\makeatletter
\def\@email#1#2{%
 \endgroup
 \patchcmd{\titleblock@produce}
  {\frontmatter@RRAPformat}
  {\frontmatter@RRAPformat{\produce@RRAP{*#1\href{mailto:#2}{#2}}}\frontmatter@RRAPformat}
  {}{}
}%
\makeatother
\begin{document}

\title[Electron Ptychography]{Electron Ptychography}

% Force line breaks with \\
\author{L. Clark}
 \affiliation{School of Physics Engineering and Technology, University of York, YO10 5DD, UK}
 \email{laura.a.clark@york.ac.uk}

\author{P. D. Nellist}%
 \affiliation{Department of Materials, University of Oxford, 16 Parks Road, Oxford OX1 3PH, UK}

\date{\today}% It is always \today, today,

\begin{abstract}
Electron ptychography describes a family of algorithms which are used to enable the reconstruction of complex specimen transmission functions of a sample in order to obtain both phase and amplitude information, as applied within the realms of electron microscopy. Ptychographic methods can be very useful in the imaging of beam sensitive materials, samples with both strongly- and weakly-scattering elements, and for mapping the functional properties of materials. Ptychography can further be used to achieve image resolutions far beyond the conventional resolution limit defined by the imaging aperture size.
In this article, we review the development of ptychography and compare presently-available methods to perform ptychographic image reconstruction on data collected in the (scanning) (transmission) electron microscope. We aim to enable the reader to determine optimal data collection parameters, and the most suitable ptychographic method for the information they seek to determine from their sample.
\end{abstract}

\maketitle

\section{\label{sec:intro}Introduction}
\epigraph{On the Roman Wall, I have omitted much that I would have liked to have said... On the Roman Walls, I have written much that few will read.}{John Hodgson \citep{hodgson}}
%[\textipa{ta\textsci k\textturnscripta  g\textturnr \textschwa fi}:]
Ptychography \href{http://ipa-reader.xyz/?text=ta%C9%AAk%C9%92g%C9%B9%C9%99fi%3A&voice=Brian}{[\textipa{ta\textsci 'k\textturnscripta  g\textturnr \textschwa fi}:]}
describes a loosely grouped family of computational imaging algorithms which allow us to measure the full complex form of the specimen transmission function(s) of a sample, through multiple illumination conditions. In this way we can overcome the \emph{phase problem} as common to many imaging sciences: the record-able data contains only amplitude information of the complex function of interest and one must devise strategies in order to retrieve the phase  of that complex function.
In electron microscopy, the phase of the specimen transmission function  contains information key to understanding specimen properties, such as electric or magnetic fields, or the presence of light (low atomic number) elements - with further complexity in many situations due to multiple or dynamical scattering \cite{MacGillavry1940prufung}.

Conventional electron microscopy imaging techniques, such as bright-field transmission electron microscopy (BF-TEM), or high-angle annular dark field (HAADF) scanning-TEM (STEM),  do not require computational processing after acquiring a 2D image, but do not provide access to the full complex specimen transmission function which is often sought \cite{WilliamsCarter, hawkes2019springer}. Computational imaging methods, such as ptychography, require one or more computational processing steps between data-acquisition and image-generation. It is through this computational processing that one can enable access to the complex specimen information.

There are many ways to approach the phase problem in the imaging sciences and we might broadly categorise these into coherent diffractive imaging (CDI) approaches, holography or ptychography \cite{hawkes_spence_2007}. In CDI, a sample is illuminated in one optical plane, and an intensity recording is made in another optical plane. This is most often in the far-field diffraction plane and as such the recorded data is the intensity of the Fourier transform of the product of the specimen transmission function and the illumination profile, but near-field approaches also exist. The uncountably-many potential specimen transmission functions that would form any given recorded intensity profiles in the detector plane can then be restricted by applying physically-meaningful constraints, such as specimen atomicity \cite{sayre1952squaring}, analyticity \cite{gerchberg1974super}, or finite extent \cite{fienup1987reconstruction}. A comparison of some of these approaches found in Ref. \cite{fienup1982phase}. These tools are commonly exploited in X-ray imaging and crystallography \cite{miao2011coherent, robinson2009coherent, miao2015beyond, sayre1995x}.

Holography imposes different constraint - one records the interference pattern between a known reference wave and the wave of interest \cite{gaborholog}. The desired complex specimen information is then retrieved from the recorded hologram intensity through a Fourier filtering process (in the typical setup of off-axis holography, performed experimentally using a charged wire as a biprism to form a tilted plane wave reference, one can find the object information laterally displaced from the zero-order information) \cite{hawkes2019springer}.

Alternatively in ptychography, one records an intensity pattern (typically a far-field diffraction pattern), moves the illumination (relative to the sample), and records another intensity pattern. In this way, the dataset is constrained by \emph{a priori} knowledge of the illumination and permits solution through a range of algorithmic approaches.

We argue here, that ptychography provides a more flexible experimental approach than CDI or holography, and that there exists an experimental design and ptychographic algorithm to suit a great many situations of interest to the electron microscopist. Ptychography needs only a segmented or pixellated detector, and can be performed using plane waves (conventional transmission electron microscopy (TEM)-like illumination), with a scanned focussed probe (scanning TEM (STEM)-like) or intermediate optical settings \cite{clark2023effect, martinez2019direct, haigh2009atomic, ClarkFourierPtych}.

The name ``ptychography'' was first coined in  1970 \citep{hegerl1970dynamische} and is used to describe an approach to overcome the phase problem through collecting multiple data points while varying the illumination, with the algorithms  enabling an ``unfolding", from the Greek, $\pi \tau \upsilon \chi \acute{\eta}$, meaning \emph{to fold} \citep{hegerl1970dynamische}, via the German \emph{Faltung}, meaning both folding and convolution. As such we consider ptychography as \emph{deconvolution}-enabled imaging.
Further methods and experimental demonstrations followed through the late 1900s and into the early $\mathrm{21^{st}}$ century and will be discussed in this article.
  
Ptychographic algorithms are used across microscopy and imaging sciences to improve image resolution, dose-efficiency and contrast when compared to conventional imaging methods. These methods are used across much of the electromagnetic spectrum, from X-ray to ultraviolet to visible light, and across a broad range of electron energies (from 30--300keV), for a vast array of applications \cite{pfeiffer2018x, odstrcil2015ptychographic, maiden2010optical, humphry2012ptychographic, chen2021electron, maiden2009improved, muller2014atomic}.
In this review article, we will restrict ourselves  to the discussion of ptychography as used in the electron microscope (and primarily TEM and STEM, while noting the work applied in SEMs), giving an overview of the main methods and their development along with highlights of key applications enabled by these tools.

This article will proceed by firstly describing a brief history of the methodology in Section \ref{sec:history}, before an overview of how imaging processes are scaffolded by the Fourier transform, along with 4D-data structure fundamentals in Section \ref{sec:datastructure} before  describing the key algorithms used in the community in Section \ref{sec:algorithms}, their applications and limitations (Section \ref{sec:applications})  before taking a broader view of how these methods are related to allied imaging tools and techniques in Section \ref{sec:Conc}. Appendices are included to provide a shorthand access to definitions \ref{sec:definitions} and accessible analysis codes \ref{sec:codes}.

\section{\label{sec:history}A Historical View of Ptychography}

The phase problem is a long standing challenge across the imaging sciences \cite{taylor2003phase, misell2023phase}.
An early solution was suggested by Gabor in 1947 \cite{gabor1971holography, gaborholog}, via holographic reconstruction. This was impractical with the instruments of the time, but has developed significantly in the subsequent decades \cite{hawkes2019springer}. Nowadays, electron holography  is experimentally feasible on a number of instruments, but faces constraints around needing a nearby vacuum region and inelastic and multiple scattering. Electron ptychography presents an alternative solution to this problem, and we suggest is nowadays experimentally simple to perform. Herein we review the development of this field, before presenting current methods to solve for the phase through a ptychographic approach.

From the early days of TEM and STEM imaging, it has been known that the detectors were only able to collect part of the signal impinging on them - the inability to measure phase and amplitude simultaneously is a fundamental limitation of measuring quantum mechanical wavefunctions.
The earliest suggestion we have found of a method to measure phase in a ptychographic manner, is in an aside from  \citet{menter1956direct} in 1956:
\begin{quote}
"In the course of a discussion with Dr P. H. [sic]  Hirsch it was realized that it may be
possible to use the effects described in this paper for making a rough evaluation of
the relative phases of all the diffracted beams and so make a Fourier synthesis of
a crystal structure by much simpler methods than those conventionally used in
structure determination by diffraction."
\end{quote}

The evaluation of the relative phases of diffracted beams referred to by Hirsch would enable the wave in the diffraction plane to be characterised in terms of both phase and amplitude (in this case, by image formation using a narrow aperture, selecting different triads of scattered beams in turn), in order to enable a full measurement of the complex specimen transmission function and a full characterisation of the crystal structure. Such an experiment was infeasible at the time, due to the high-demands placed on microscope stability. As such, the first detailed experimental proposals and subsequent data followed slowly in the subsequent decades.

\begin{figure}[!ht]
  \centering
  \includegraphics[width=.7\linewidth]{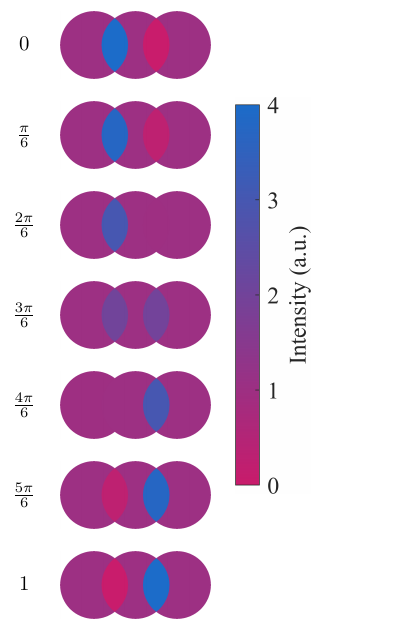}
  \caption{Schematic illustration of coherent interference as described by Hoppe in 1969 \citep{Hoppe_1969_I}. Here we see diffraction patterns from a sinusoidal grating with a round condenser aperture. The central disc represents the unscattered beam, with the side discs being the scattered beams (left and right beams are complex conjugates of each other and thus always $\pi$ out of phase with each other). When the sample imposes a $\pi$ phase shift on one of the scattered beams, we see constructive interference in one overlap region and destructive interference in the other - leading to differences in measured intensities (top row). When the sample imposes other phase shifts on the scattered beams, the relative intensities in the two overlap regions will change, in proportion to the sample imposed phase shift, and measurable on a simple diffraction pattern.}
  \label{tab:interference}
  \end{figure}

Hoppe \citep{Hoppe_1969_I, Hoppe_1969_II} developed the idea of unfolding complex wave information from interference effects through this approach, between interference of diffraction spots as a result of the objective aperture. 
This approach was broadened to encompass non-crystalline specimens in Ref. \cite{Hoppe_1969_III}. A sufficiently small aperture will lead to these discs overlapping - and as this is a coherent imaging process - in the overlap region, the wavefunctions will interfere coherently and the recorded intensities of these interference patterns will reveal their relative phase shifts, as illustrated in Fig. \ref{tab:interference}.

By the start of the 1970s, it was acknowledged that an experimentally feasible approach for the instruments of the time would do well to make use of the flexible optics available in the TEM: `` One can make several exposures under different conditions, provided that in the combination of the diagrams the complete information can be found'' \cite{hoppe1970principles}.

Through the 1970s and 80s, there were a number of advances made in CDI approaches, which later fed into ptychographic methods.
In a related approach, Gerchberg and Saxton published an iterative phase retrieval approach in 1971 \cite{gerchberg1971phase, gerchberg1972practical}, and almost simultaneously, a patent was submitted containing an iterative phase retrieval method by Hirsch \emph{et al.} \cite{hirsch1971method}.

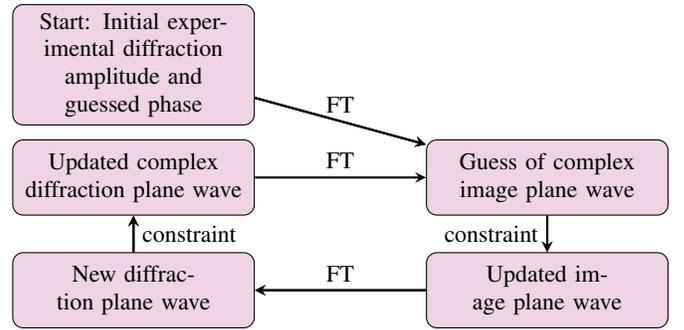
\begin{figure}[!ht]
\centering
\begin{tikzpicture}[node distance=1.5cm]
\node (aa) [GSbox] {Updated complex diffraction plane wave};
\node (bb) [GSbox, right of=aa, xshift=4cm] {Guess of complex image plane wave};
\node (cc) [GSbox, below of=bb] {Updated image plane wave};
\node (dd) [GSbox, below of=aa] {New diffraction plane wave};
\node (ee) [GSbox, above of=aa] {Start: Initial experimental diffraction amplitude and guessed phase};
\draw [arrow] (bb) -- (cc);
\draw [arrow] (cc) -- (dd);
\draw [arrow] (dd) -- (aa);
\draw [arrow] (ee) -- (bb);
\draw [arrow] (ee) -- node[anchor=south] {FT} (bb);
\draw [arrow] (aa) -- node[anchor=south] {FT} (bb);
\draw [arrow] (cc) -- node[anchor=south] {FT} (dd);
\draw [arrow] (bb) -- node[anchor=east] {constraint} (cc);
\draw [arrow] (dd) -- node[anchor=west] {constraint} (aa);
\end{tikzpicture}
\caption{Schematic of Gerchberg-Saxton phase retrieval approach as in selected area (SA)-TEM, with a recorded diffraction pattern and apertured-image pair. The image-plane constraint would be to force amplitude to zero outside of the known aperture area, and the diffraction-plane constraint would be to enforce amplitudes from the square-root of the recorded diffraction intensity.}
\label{fig:GSalg}
\end{figure}

The Gerchberg-Saxton method uses the \emph{a priori} knowledge (i.e. knowing the modulus of the wave in both real and reciprocal space) to iteratively constrain the solution until a sufficiently close match is attained. Their  approach, illustrated in Fig. \ref{fig:GSalg}, underlies much of the subsequent work in iterative phasing.  If one compares the Gerchberg-Saxton method with the subsequent iterative ptychographic approaches (discussed later in this article), the parallels are apparent. Subsequent work by Fienup modified the approach of Gerchberg and Saxton such that the real-space constraint is loosened to a requirement for non-negativity and object support \cite{fienup1982phase, elser2003phase} which can be more suitable for some experimental setups.

In the following years, this approach was developed further. Gerchberg \cite{gerchberg1974super} suggested a route towards superresolution, while Liu and Gallagher further developed understanding of the error-behaviours \cite{liu1974convergence}. Fienup's later development of the hybrid input-output method enabled slower-but-stabler convergence behaviours \cite{fienup1987reconstruction, fienup2013phase}, and noted that iterative methods can themselves be iterated between to improve the overall solution \cite{fienup1978reconstruction}.

By the late 1980s and into the 1990s, approaches were also made toward direct methods for solving the phase problem through a ptychographic approach by Rodenburg and collaborators \cite{bates1989sub, mccallum1992two, friedman1992optical, mccallum1995complex, nellist1995resolution, rodenburg1993experimental}. In particular, we highlight here the major developments made by Rodenburg and Bates \cite{rodenburg1992theory}, building underlying scaffolding  for the understanding of four-dimensional information transfer processes - including insights to finite coherence, finite dose, and impact of microscope aberrations.

All of which together pave the way for the algorithms currently used, and will be explored in more detail in the sections below.

As an aside here, we also note an alternative approach to sample structure determination: solving for the scattering matrix, by employing 4D-STEM datasets. This S-matrix approach was developed around the turn of the millenium \cite{allen1999retrieval, allen2000inversion} with subsequent significant advances made by Findlay and collaborators  \cite{findlay2005quantitative, brown2018structure}. This methodology faces similar challenges to our ptychographic approaches \cite{sadri2023determining} and comparisons between the two methods may prove insightful.

Ptychography is often considered to be a very distinct imaging process from other, more direct, imaging methods in electron microscopy - but this really needn't be the case. In 1982, Hawkes \cite{hawkes1982stem} highlighted that the STEM is an ideal instrument for ptychography, questioning if all STEM-users were indeed ``unwitting ptychographers''. The interference patterns that we make use of to analyse in our STEM ptychography, are always there in the experiments - and can be detected as long as a sufficiently finely segmented detector is used. The segmented (quadrant) detectors as used in DPC-STEM to determine an approximation of the centre-of-mass of the bright-field disc are indeed a good first-order approximation to the data needed for ptychographic imaging \cite{rose1976nonstandard, chapman1992differential, shibatafindlay}. Three detectors are sufficient \cite{mccallum1995complex,brown2016structure}, though four might be easier to interpret in a Cartesian co-ordinate system. The 16 segments available in the segmented annular detector provide plenty of data to approximate the key Fourier components contained in the transmitted data \cite{shibata2010new}. Indeed, some preliminary studies suggest that of the order of 100 pixels may suffice for specific experimental settings \cite{yang2015efficient, zhang2021many, muller2017measurement}.

\section{\label{sec:datastructure}Geometries and data structures underlying ptychography}
Before we proceed into a detailed discussion of currently used ptychographic algorithms and their applications, we will first run through some fundamentals of 4D datasets and their collection to orient the reader.

\subsection{\label{sec:Fourier}Imaging as applied Fourier transforms}
In order to understand ptychography (that is, convolution-enabled imaging), one must understand how the Fourier transform operates on data between real and reciprocal space, and how manipulations on that data in one space will impact the data available in its conjugate space. In this section, we aim to remind the reader of some of the key properties of Fourier analysis as pertinent to ptychographic theory.

Fourier transforms are crucially important to both image formation and image analysis. In the image-formation process in a model scanning transmission electron microscope, the wavefront impinging on the condenser aperture plane is Fourier transformed by a lens such that the circular illumination passed through the aperture is transformed to an Airy disc illumination profile describing the probe shape on the sample (in the far-field, or back focal plane of the condenser lens). If the sample is reasonably thin, the majority of this illuminating beam passes through the sample and is then Fourier transformed once again by another lens, resulting in a circular disc of electron intensity impinging on the detector below. Structure and patterning in this detected circular disc of intensity reveal interactions of the sample with the transmitted electron beam.

Image analysis is also substantially composed of operations on the Fourier transform of an the image - smoothing and noise reduction is typically achieved by Fourier transforming an image, applying a low-pass filter, and then inverse Fourier transforming the dataset. The resolution achieved in an electron microscopy image is often defined by Fourier transforming a recorded image, and finding the position of the highest scattering-angle at which there is observable feature (which translates to the shortest-distance periodicity in the real space image).

If we could directly record the full complex wavefunction transmitted through the sample ($\Psi=|\Psi|\mathrm{exp}(\mathrm{i}\phi)$), there would be little utility in ptychographic algorithms - however, we can currently only detect the total electron intensity ($|\Psi|^2$)  so we must make use of experimental and mathematical tools to extract the phase information we seek from the data that can be experimentally recorded. The key experimental tool of note for STEM ptychography is use of a segmented or pixellated electron detector, which enables recording the position of each electron impinging on the specimen. These detectors are increasingly widely available and are further discussed in Ref. \cite{ophus2019four}. For the recorded data to be most useful, it is also helpful to ensure the electron beam is as coherent as possible as this enables interference effects to be recorded clearly - coherence properties are a function of both the type of electron source and the optical settings.

\subsubsection{Significance of accurate phase for accurate interpretation}
\begin{figure}[!ht]
  \centering
  \subfloat[][]{\includegraphics[width=.4\linewidth]{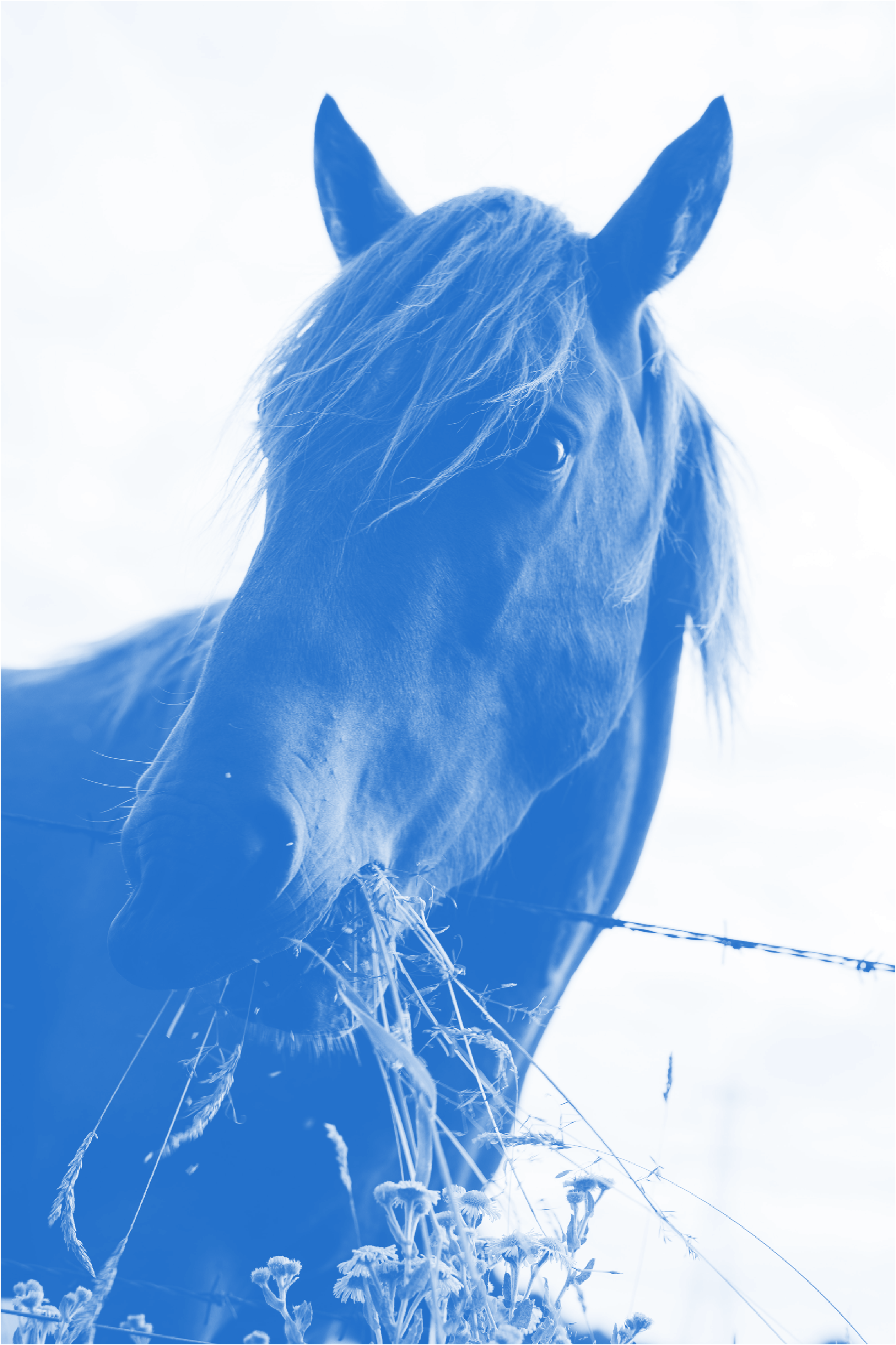}}\quad
  \subfloat[][]{\includegraphics[width=.4\linewidth]{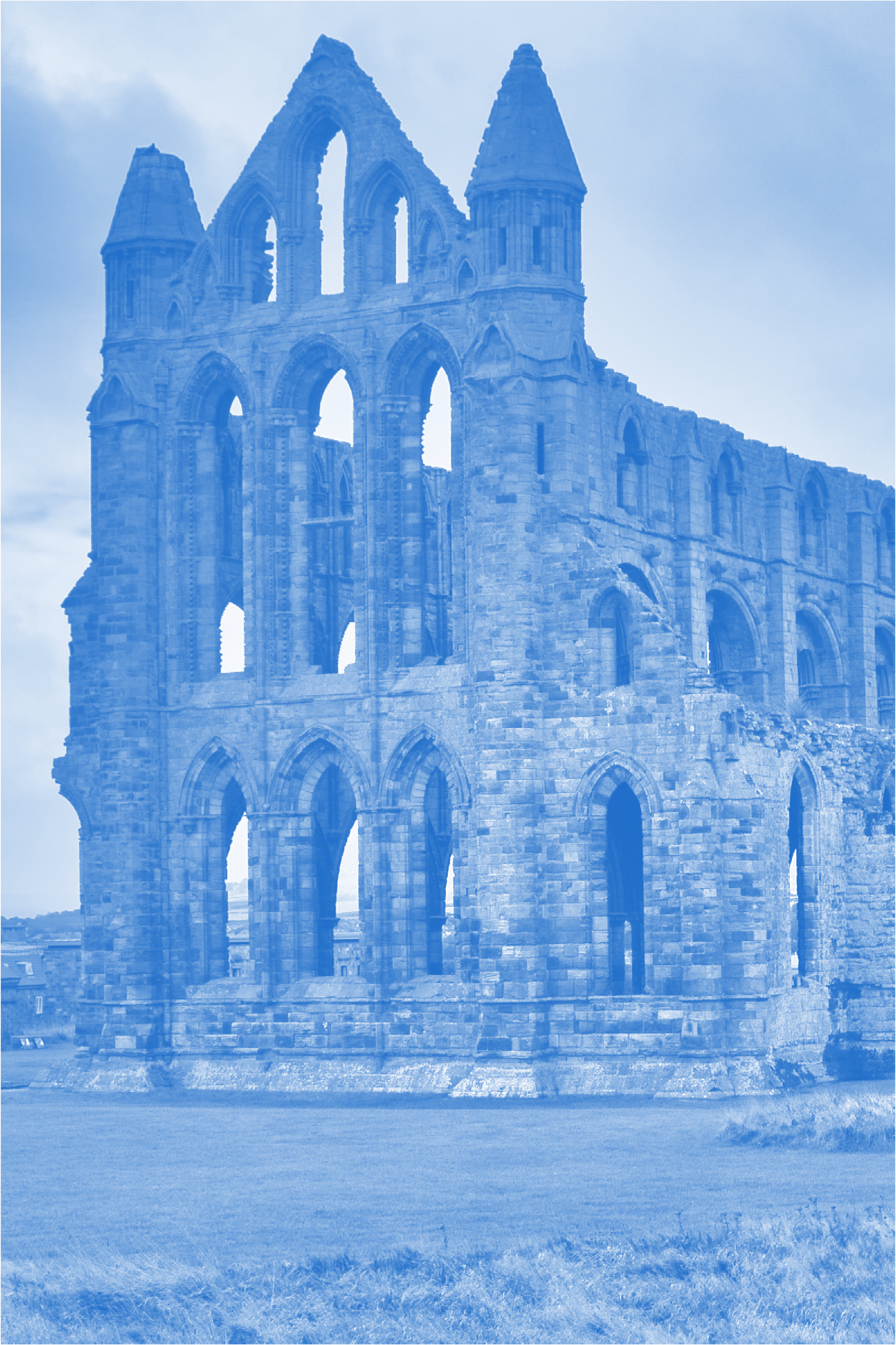}}\\
  \subfloat[][]{\includegraphics[width=.4\linewidth]{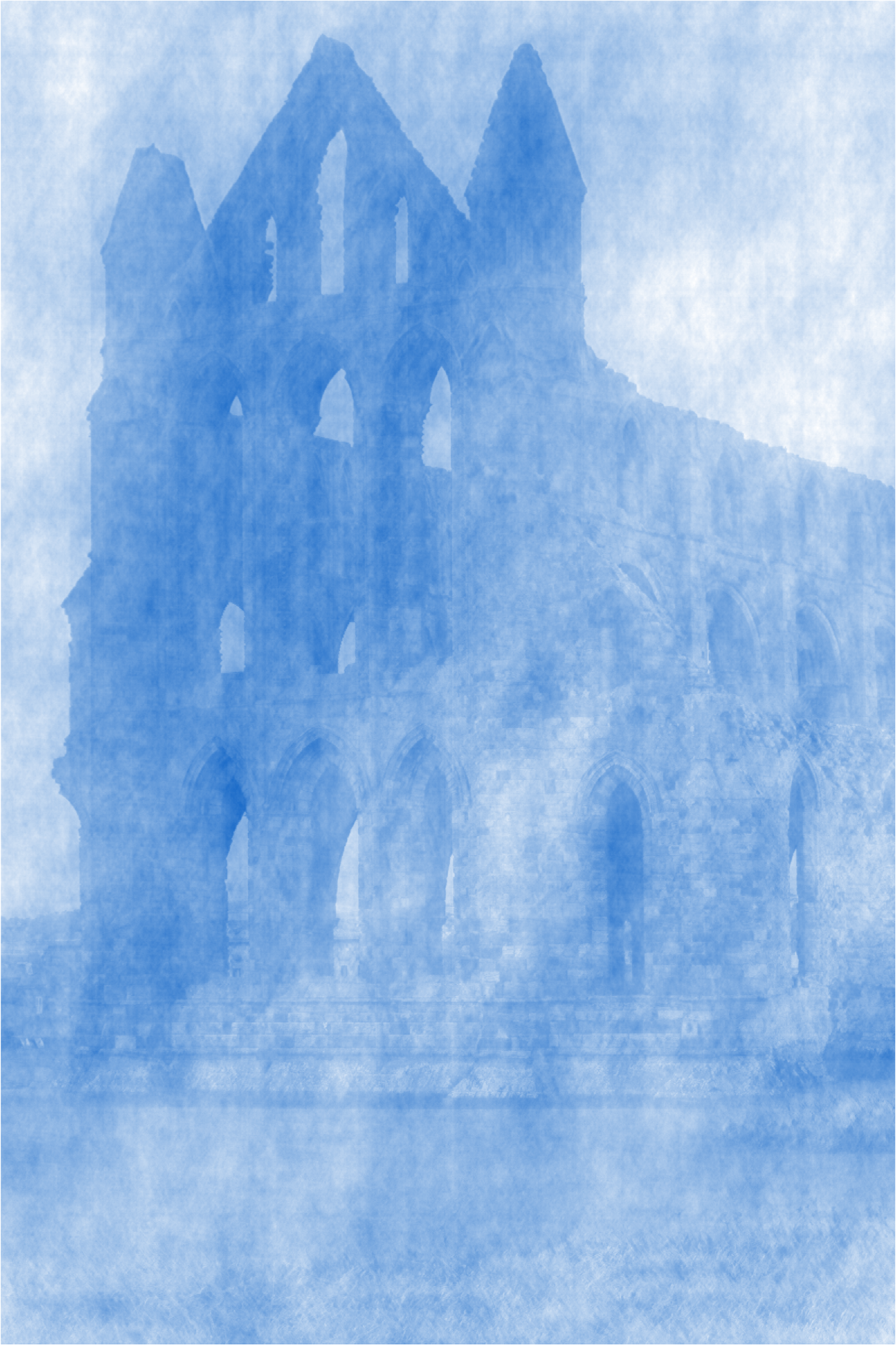}}\quad
  \subfloat[][]{\includegraphics[width=.4\linewidth]{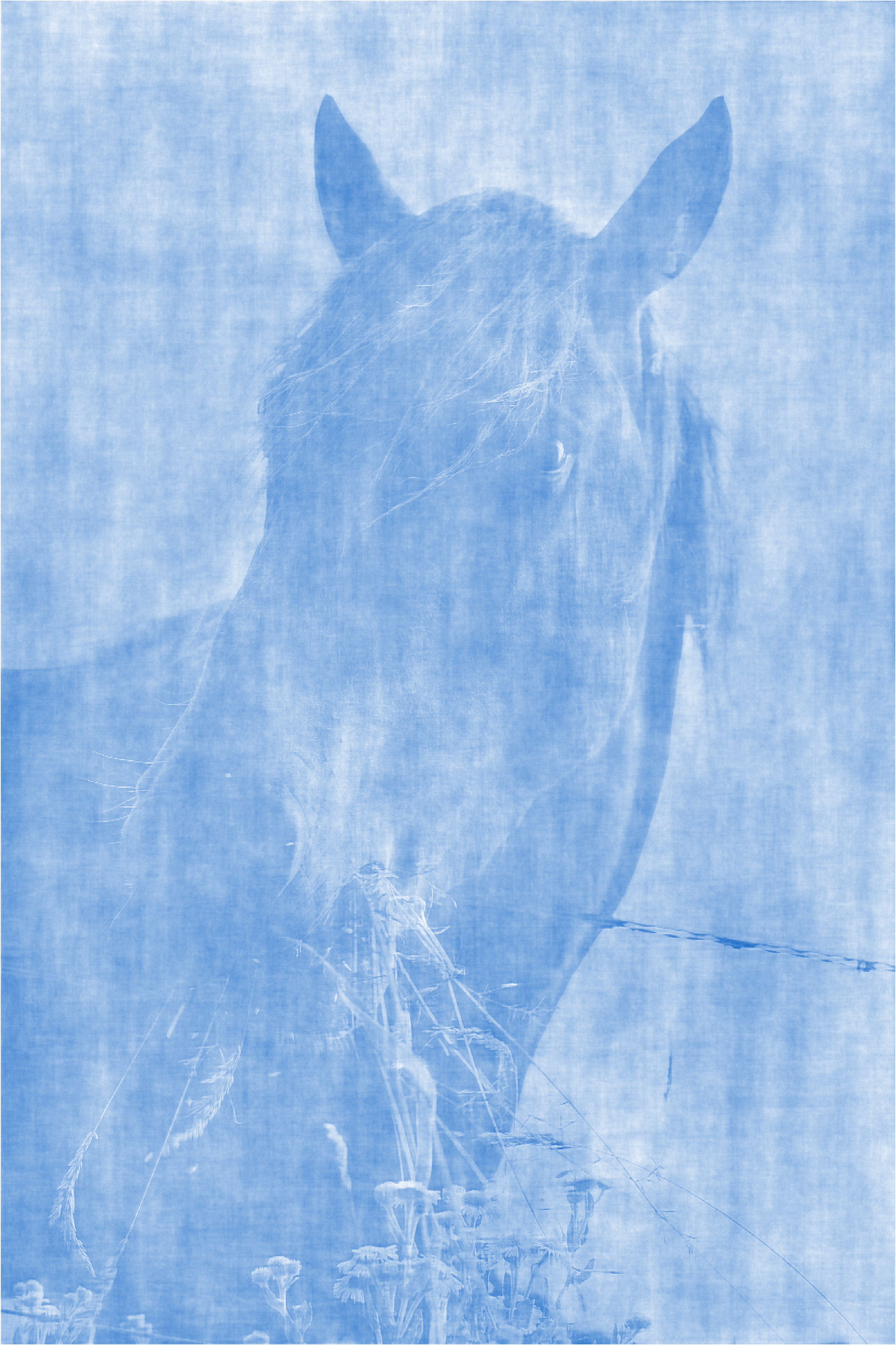}}
  \caption{(a, b) Real-valued example images of an inquisitive horse and Whitby abbey, respectively, (c, d) The real part of complex images resulting from phase-swapping in Fourier space: (c) Result of Fourier amplitude of \emph{(a)} with Fourier phase of \emph{(b)}, (d) Result of Fourier amplitude of \emph{(b)} with Fourier phase of \emph{(a)}. Input figure data courtesy of Refs. \cite{horse, abbey}.}
  \label{fig:phaseswap}
  \end{figure}
If we have an arbitrarily-structured real-valued 2D array of data (such as the two images in Figure \ref{fig:phaseswap} (a) and (b), or equally any recorded conventional image from the electron microscope), we can perform a 2D Fourier transform on that data and find we have a 2D array of complex values. If we retain that complex information and perform an inverse Fourier transform, we get back to the initial dataset. However, if we were to mis-handle the phase structure of that complex data (for example, swapping the phases of the Fourier transforms of Figure \ref{fig:phaseswap}(a) and (b)) before performing the inverse Fourier transform, we find that the image no longer reconstructs as expected (illustrated in Figure \ref{fig:phaseswap} (c) and (d)). We find that (c) more closely resembles (b) and likewise (d) more closely resembles (a): the phase component in the Fourier transform of the data is crucial to form an reliable image. 
This is because the phase structure (and more specifically, the local gradients of that phase structure) in the Fourier plane determines the positions of image features in real space, and we (as human interpreters) primarily interpret images by the relative positions of the different features. This scrambling of features by phase modification is employed by neuroscience researchers to segregate brain signals due image intensity variations from signals due to image feature interpretation \cite{willenbockel2010controlling}. Similarly in electron microscopy studies, the relative postions of sample features is usually key information, and as such the phase structure of the data is fundamental to a successful microscopic investigation. These concepts are expanded more formally below.

\subsubsection{Key Fourier analysis theorems}

There are two specific Fourier analysis theorems we feel are key to understanding the operations of ptychographic imaging. If our real space data (a 2D real or complex array of values \footnote{We will reserve discussion of 3D structures until \ref{sec:multislice}}) is expressible as $f(x,y)$, we can describe its Fourier transform as:
\begin{equation}
    \mathcal{F}(f(x,y))=F(k_x, k_y).
\end{equation}
$F(k_x, k_y)$ will also be a 2D array of values (which will be real, imaginary or complex depending on the input symmetries and form of $f$). In the general case, $f(x,y)$ is complex, as is its Fourier transform, $F(k_x, k_y)$.

The Fourier shift theorem (e.g. Eq. 2.13 in Ref. \cite{goodman2005introduction}) states that a lateral rigid shift (of $(x_a, y_b)$ ) of the wavefunction in one space will lead to a linear phase gradient being applied to its Fourier conjugate:
\begin{equation}
    \mathcal{F}(f(x-x_a,y-y_b))=F(k_x, k_y)\cdot \mathrm{exp}(\mathrm{2\pi i (k_x a +k_y b)}),
    \label{eq:shift}
\end{equation}
where $a$ and $b$ are arbitrary real numbers, with $x_a,y_b$ their reciprocal coordinates.Incidentally, this theorem is the key concept underlying differential phase contrast (DPC) STEM \cite{rose1976nonstandard, clark2018probing}, a non-ptychographic method of phase retrieval from a 4D-STEM dataset, wherein a phase gradient at the sample plane results in an intensity shift in the detector plane.

Eq. \ref{eq:shift} is important because it emphasises that without phase information in one space we cannot determine the relative position of features in the conjugate space. This is precisely why the Patterson function (as in Refs. \cite{patterson1934fourier} and \cite[Chapter~12]{sherwood2010crystals}) can tell us  spacing between features in a specimen, but not the position, or relative positions of features (without additional filtering and restrictions, such as applied by a multi-pinhole interferometer \cite{clark2014quantitative}).

The second key theorem underlying ptychographic analyses is the convolution theorem. The data we record in the microscope is the data we want, convolved with microscope parameters we are typically not particularly interested in. Approaches to deconvolve this data often employ the result that one can describe the convolution of two functions as the Fourier transform of the two signals multiplied together. Multiplications and divisions are much easier to filter apart. More formally (following Eq. 2.15 of Ref. \citep{goodman2005introduction}) we express this theorem as:
\begin{equation}
    A \circledast B = \mathcal{F}(a \times b).
\end{equation}
Where $A=\mathcal{F}(a)$, $B=\mathcal{F}(b)$ and $\circledast$ denotes convolution. This concept is particularly crucial in locating the signal in the double overlap region of single side-band ptychography (discussed further in section \ref{sec:algorithms}).

\subsection{\label{sec:datasets}Four dimensional datasets and their acquisition}
 The four dimensional datasets required for ptychographic analyses in (S)(T)EM, can be acquired in three broad types: (i) focused-probe STEM, (ii) defocused-probe STEM or (iii) tilt-defocus TEM (which makes dataset with between 4 and 5D depending on which way you look at it).  In this section, we discuss the nomenclature used to describe these datasets, the detectors used to acquire these datasets and a generalised description of these datasets, as used in the algorithms discussed in the subsequent section \ref{sec:algorithms}.

\subsection{Nomenclature of four dimensional STEM}
There are a variety of terms used in the community to describe similar and overlapping imaging techniques - such as scanning electron nanodiffraction (SEND), electron microdiffraction, nanobeam electron diffraction (NBED) and momentum-resolved STEM or diffractive imaging.
In this article, we use the phrase `four-dimensional STEM' (4D-STEM) as an overarching descriptor for these methodologies in which an electron probe is raster scanned across a 2D array of probe positions, and at each probe position a 2D array of electron intensities is collected.

\subsection{Detectors for 4D-STEM}
In order to record 4D-STEM data, one must have a STEM detector with multiple pixels. In earlier work, this was typically performed through careful use of a pixelated TEM detector - with associated challenges of slow camera speed, sample drift and potential for camera damage \cite{nellist1995resolution}. An alternative route forwards is to use the segmented and segmented-annular detectors developed through the Lorentz microscopy community \cite{dekkers1974differential, rose1976nonstandard, seki2018theoretical}, as indeed, ptychographic reconstruction is as possible with as few as three detector segments \cite{mccallum1995complex, brown2016structure}\footnote{The evolution of multiple detectors is described by \citet{kirkland2020advanced} and the details of nomenclature choice are discussed in more detail by \citet{ophus2019four}.}. For higher resolution and improved sample sensitivity however, more pixels are beneficial. 

4D-STEM for ptychographic reconstruction can more easily be performed however, by taking advantage of recent developments in fast direct electron detectors \cite{faruqi2018direct, paterson2020fast, philipp2022very}, which allow a dataframe of thousands of pixels to be collected at each probe position in the STEM raster scan.
The question then becomes one of how to balance data-size (10GB per STEM scan is not uncommon) with information obtained from the specimen. Options to consider here would include cropping or binning the data before further processing \cite{savitzky2021py4dstem}, or reducing the bit-depth of the data \cite{o2020phase} - in addition to careful choice of field of view, number of probe positions and real space probe-step-size.

In the remainder of this manuscript, we will consider 4D-STEM datasets to be those collected on pixellated datasets on a Cartesian array ($N \times N$ pixels, where $N$ is a reasonably large number.)

    \begin{figure}
    \centering
    \includegraphics[width=0.7\linewidth]{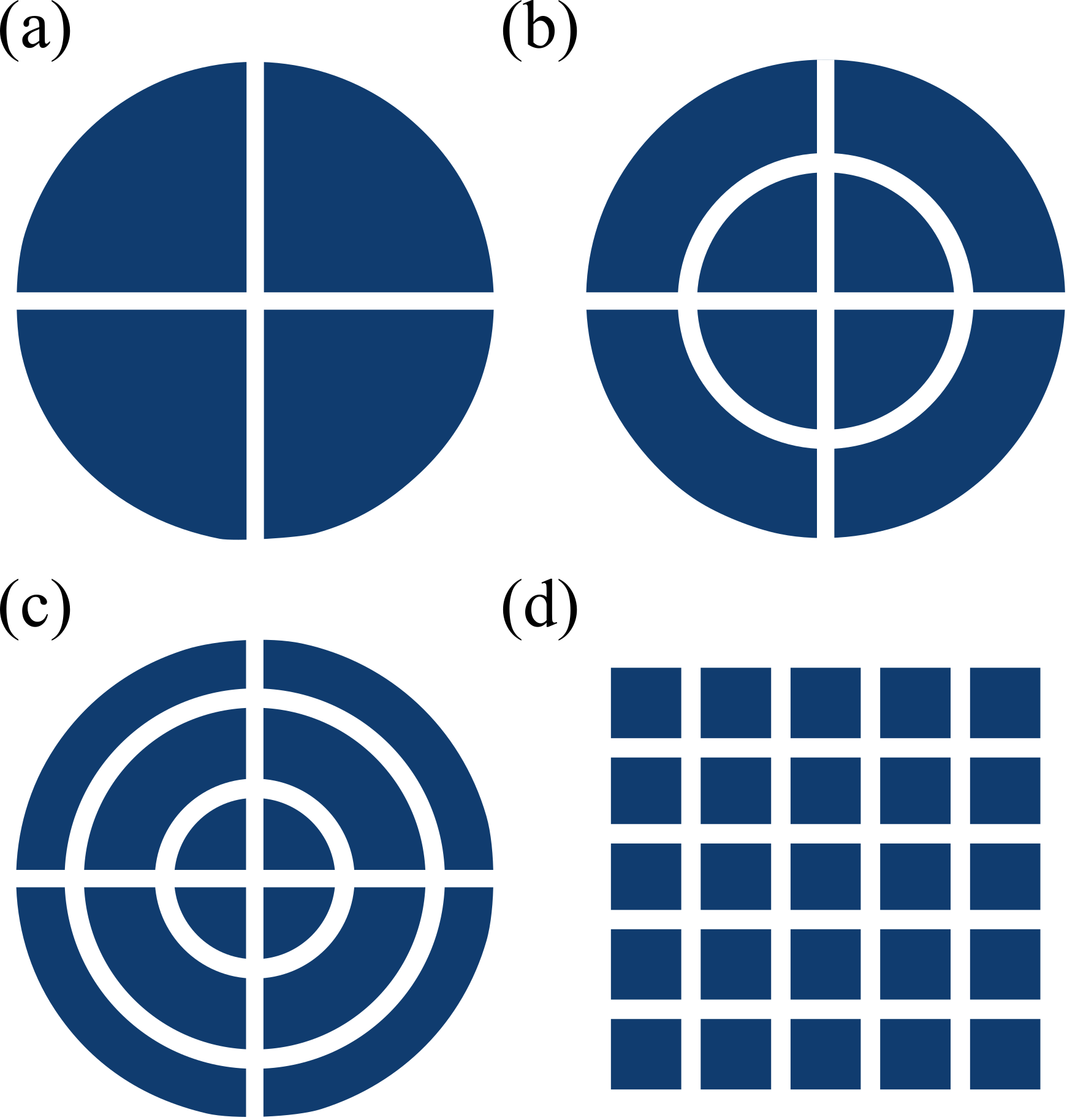}
    \caption{Schematic illustration of segmented detector geometries: (a) a conventional bright-field detector divided into four sectors will allow for an approximate measurement of specimen phase gradient in two orthogonal directions \citep{dekkers1974differential, rose1976nonstandard}, (b) with the further radial divide, one can gain a signal-to-noise boost in some cases \cite{chapman1990modified}, (c) further dividing the detector into a segmented-annular case with 16 shaped pixels improved SNR and better reciprocal space sampling \citep{AnnularSegmented}, (d) a detector made of a Cartesian grid of many independent pixels allows for the greatest sensitivity to scattering angle.}
    \label{fig:detectors}
\end{figure}

\subsubsection{Fast detectors and live processing}

While electron ptychography was first successfully demonstrated experimentally in the  1990s \cite{nellist1995resolution}, with some subsequent exploratory work \cite{nellist1998electron, plamann1998electron},
developments in this field did not accelerate significantly until post-2010 as fast pixelated detectors began to become available such that 4D-STEM experiments could be achieved with a reasonable degree of experimental success\footnote{Earlier generations of slower detectors meant that sample drift often prevented collection of a clear dataset}. Segmented detectors (with 4--16 detectors) became increasingly used in the interim but typically applied to problems of DPC-STEM \cite{ishikawa2018direct, murakami2020magnetic} imaging rather than ptychographic analyses \cite{brown2016structure}. Iterative algorithms are generally not instantaneous (system and model dependent, convergence can take a few minutes to a few days or longer), but the direct methods are close to live on a standard laptop. Work is developing towards allowing approximations to the above methods to run fully live at microscope-acquisition limited speeds, as shown in recent work: Refs. \cite{weber2023live, welker2023live}.

\subsection{Generalised description of a 4D-STEM dataset}
From hereonin, we consider a pixelated detector with many pixels arranged on a Cartesian grid, for mathematical simplicity.

If one uses a pixellated detector, located in the far-field diffraction plane of the sample ($\mathbf{K}_f$) and a focussed electron probe to  raster scan across a sample in the $(x,y)$ plane in the conventional manner, one collects the intensity of the microdiffraction plane (in $(k_x, k_y)$ coordinates) at each probe position (each $\mathbf{R}_p$). This 4D-dataset can be described as $|M(\mathbf{K}_f, \mathbf{R}_p)|^2$.

If the sample can reasonably be described as a phase object, $\psi (\mathbf{R})$, then $M$ can be described as:
\begin{equation}\label{eq:Mm}
    M(\mathbf{K}_f, \mathbf{R}_p)= \int a (\mathbf{R}-\mathbf{R}_p)\psi (\mathbf{R}) \mathrm{exp} (2 \pi i \mathbf{K}_f \cdot \mathbf{R}) \mathrm{d}\mathbf{R},
\end{equation}

in which $A(\mathbf{K}_f)=|A(\mathbf{K}_f)|\mathrm{exp}(i \chi (\mathbf{K}_f))$ describes an aberrated aperture function and typically is a disc of uniform intensity, and hence $a(\mathbf{R})=\EuScript{F}(A(\mathbf{K}_f))$ describes the electron probe at the sample plane. In a conventional focused probe setup, $\lvert a(\mathbf{R})\rvert ^2$ is the Airy disc intensity pattern \cite{clark2018probing}.

Eq. \ref{eq:Mm} describes the convolution of the Fourier transform of the complex specimen function with the aperture function in the microdiffraction plane. This can alternatively (and perhaps more intuitively) be described as:
\begin{equation}
   M(\mathbf{K}_f, \mathbf{R}_p)=\EuScript{F}\left( a(\mathbf{R}-\mathbf{R}_p) \times \psi (\mathbf{R}) \right).
\end{equation}
 
If we were able to record the full complex function $M(\mathbf{K}_f, \mathbf{R}_p)$, rather than just the intensity, $|M(\mathbf{K}_f, \mathbf{R}_p)|^2$, there would be no need for ptychography as the  information we seek (the complex function, $\psi (\mathbf{R})$) is simply encoded within $M(\mathbf{K}_f, \mathbf{R}_p)$. Unfortunately, there is no currently known method to directly record this phase information and so we must make use of interferometric (or other) methods to reconstruct $\psi (\mathbf{R})$ from recordable datasets.

A powerful step towards unpicking where to locate the sought-after complex specimen information within a recorded 4D-STEM dataset can be taken by Fourier transforming the whole 4D-dataset with respect to probe position, $\mathbf{R}_p$, to give a function known as $\mathbf{G}$:
\begin{equation}
\begin{split}
    \mathbf{G}(\mathbf{K}_f,\mathbf{Q}_p) & =\\
    A(\mathbf{K}_f) & A^*(\mathbf{K}_f+\mathbf{Q}_p)  \otimes_{\mathbf{K}_f}\Psi(\mathbf{K}_f)\Psi^*(\mathbf{K}_f-\mathbf{Q}_p)\label{eq:bigG}.
    \end{split}
\end{equation}

In Eq. \ref{eq:bigG}, the newly introduced variable $\mathbf{Q}_p=(Q_{p_x}, Q_{p_y})$ is a reciprocal vector of the planar image space, $(x,y)$. As such, $\mathbf{Q}_p$ is described as the image spatial frequency and determines how fine a feature can be represented in the final image. The intermediary steps between Eqs. \ref{eq:Mm} and \ref{eq:bigG} are presented in full in Ref. \cite{rodenburg1993experimental} - though we note their use of different nomenclature.

$\mathbf{G}(\mathbf{K}_f,\mathbf{Q}_p)$ is the starting point for many of the ptychographic algorithms - it is at this point that the various approaches for unfolding the specimen signal can begin, and as such understanding how $\mathbf{G}$ holds specimen information underlies so much of ptychographic imaging.

\subsection{A note on reciprocity}

In seeking to determine the complex specimen transmission function from data recorded in the Fourier (or other optical transform) plane, it does not matter whether the chosen algorithm solves for the complex function in real or reciprocal space, as once one of these two is solved, applying the inverse transform will provide the full complex function in the reciprocal plane.

    \begin{figure}
    \centering
    \includegraphics[width=0.7\linewidth]{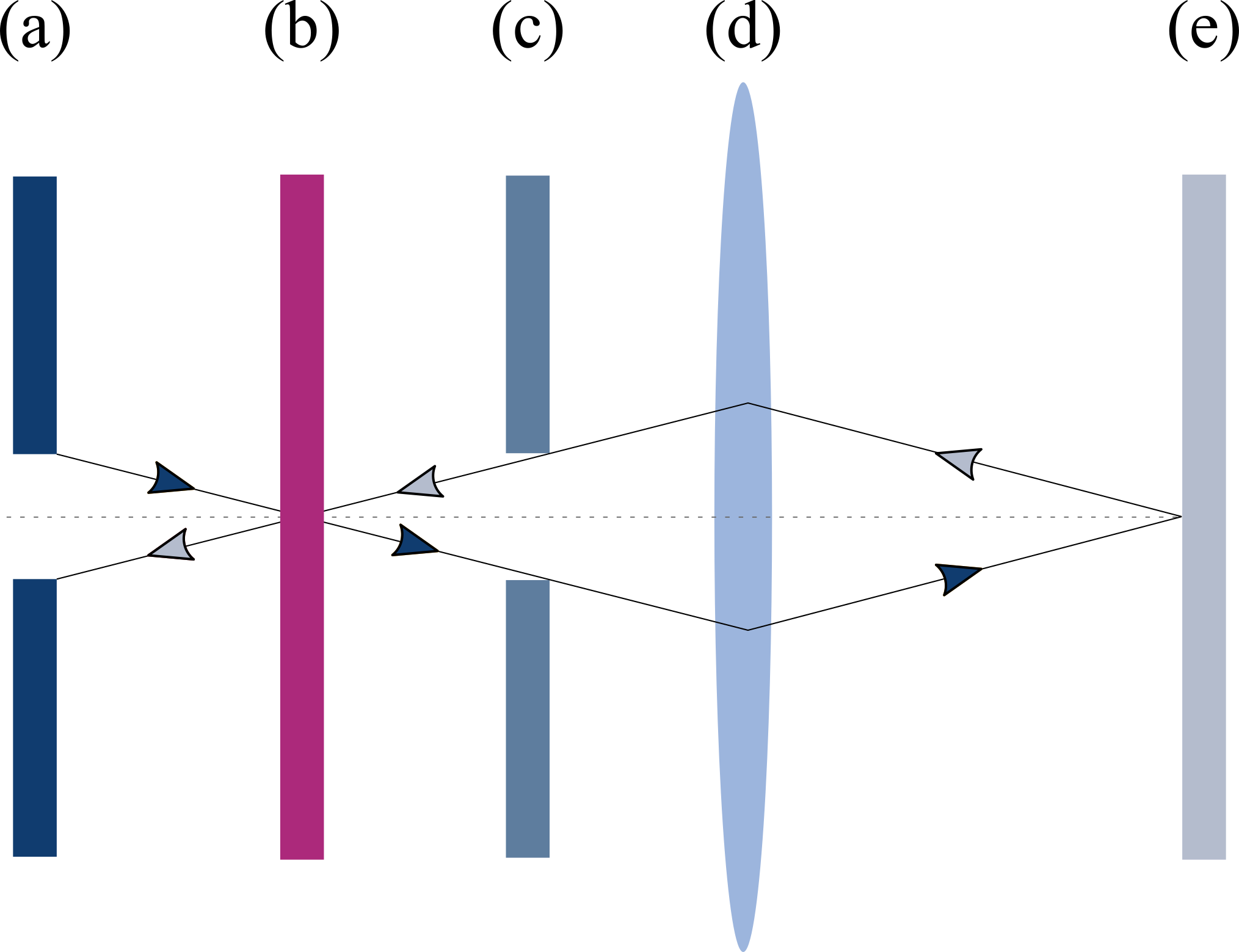}
    \caption{Illustration of the principle of reciprocity. If read left-right (dark blue arrowheads), this depicts a TEM: (a) condenser aperture, (b) specimen, (c) objective aperture, (d) objective lens, (e) detector; or if read right-left, the diagram depicts a STEM (light blue arrowheads): (e) source, (d) objective lens, (c) aperture, (b) specimen, (a) detector.}
    \label{fig:reciprocity}
\end{figure}
Many of the experimental demonstrations of electron ptychography shown in recent years are based on a STEM geometry, where an electron probe (focused or defocused) is raster scanned across the specimen.
It is perhaps illuminating to note however, that much of the theory can equivalently be applied in a conventional plane wave illumination TEM setup, with the 4D-datasets recorded across a tilt-series. It is well established that TEM and STEM datasets can be considered as Fourier reciprocals of each other as illustrated in Fig. \ref{fig:reciprocity} \citep{cowley1969image,  pogany1968reciprocity, krause2017reciprocity}, as long as we are only considering elastic scattering behaviours - this is usually the case in phase imaging experiments.

This concept is developed further below in Sec.\ref{sec:FourierPtych} on the so-called Fourier ptychography approach.

\section{\label{sec:algorithms}Ptychographic Algorithms}

To begin discussing ptychographic algorithms in some detail, we must first determine by which parameter the discussion will be ordered. We note here that this ordering is a choice of the authors, and there are many other ways in which the family of ptychographic algorithms could be arranged.
In this paper, we initially separate the algorithms by direct (\emph{i.e.} non-iterative)  or iterative, and then subdivide by features pertaining to their experimental application. This approach is chosen because of the mathematical commonalities amongst all direct methods, and separately amongst iterative methods - understanding these features are key to understanding the mechanics of the algorithms.

    \begin{figure}[!t]
    \centering
    \includegraphics[width=0.7\linewidth]{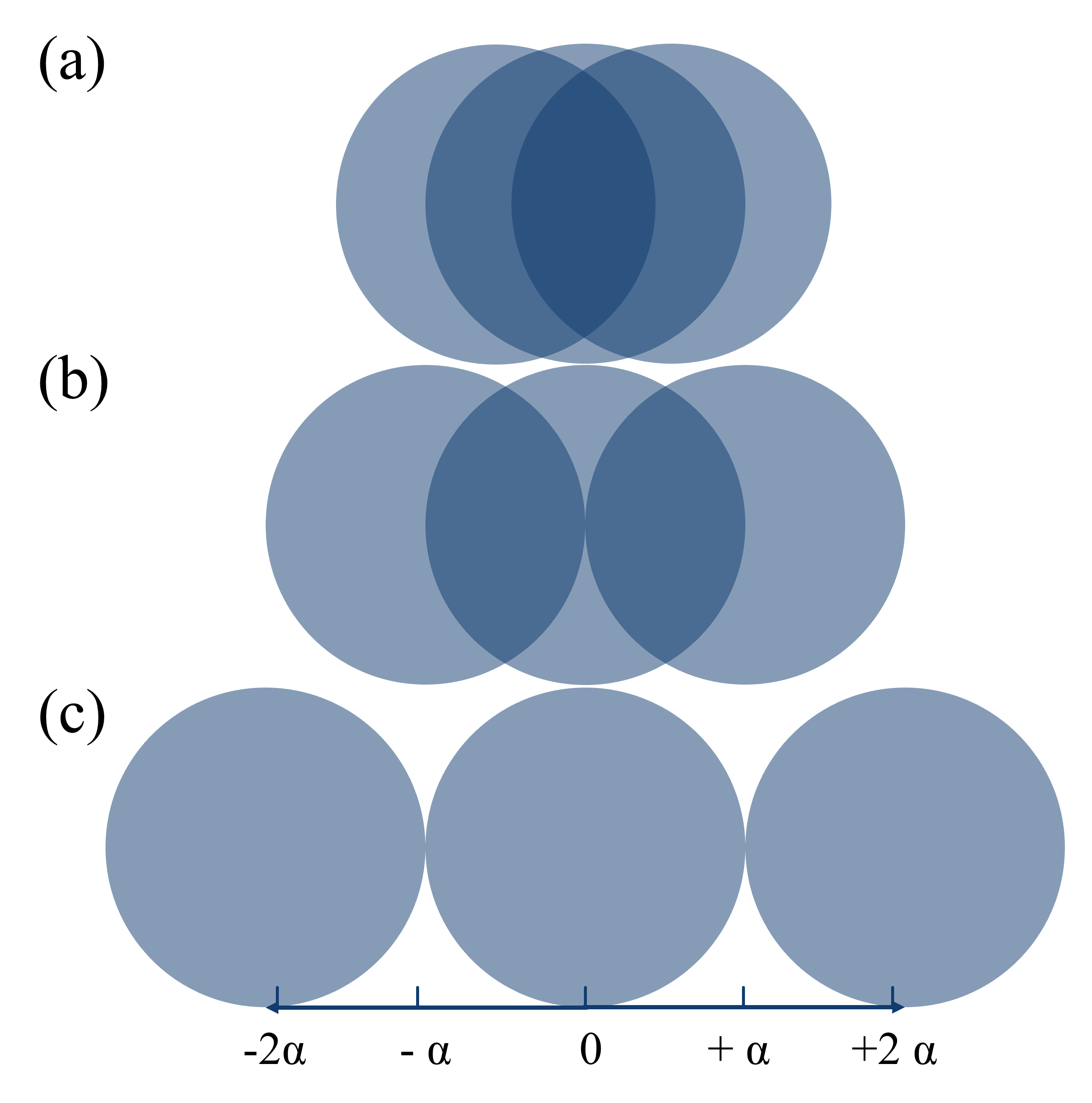}
    \caption{Schematic illustration of the overlap regions of SSB ptychography. In (a) we are inspecting a low spatial frequency, $ Q_p < \alpha$, such that the scattered side beams overlap with both the central unscattered disc, and the conjugate side beam on the opposite side, leading to a lens-shaped triple overlap region, and two double overlap regions of a more complex shape. In (b), we are inspecting an image spatial frequency, $Q_p = \alpha$, such that the region of double overlap is maximised, at which point the overlap region takes the shape of a vesica piscis \cite{fletcher2004musings}.  In (c) $Q_p = 2\alpha$, and the area of the double overlap region shrinks to zero, no contrast transfer is possible at $Q_p \geq 2\alpha$ for the non-super-resolution methods.
    }
    \label{fig:overlap}
\end{figure}
\subsection{Direct methods}
There are a range of direct (\emph{i.e.} non-iterative) methods possible to extract specimen phase information from 4D-STEM datasets. The simplest approaches include differential phase contrast (DPC)-STEM, or centre-of-mass (COM) imaging \cite{rose1976nonstandard, dekkers1974differential, clark2018probing, clark2023effect} (occasionally referred to as momentum-resolved STEM \cite{muller2018atomic}). These methods do not fundamentally rely on deconvolution processes, and so we do not consider these further here - other than to note, deconvolution processes can be applied to reduce measured aberrations in these datasets, entirely analogously to ptychographic methods \cite{ding2023}.

Direct methods of ptychography approach the phase problem through solving a finite number of simultaneous equations directly, given sufficient prior information. If sufficient prior information is available, the equations are solvable. A number of methods exist, and the primary approaches are discussed below.

\subsubsection{Single side band (SSB) ptychography}

\begin{figure}[!ht]
\centering
\begin{tikzpicture}[node distance=1.75cm]
\node (start) [startstop] {Input: $\mathbf{G}(\mathbf{K}_f,\mathbf{Q}_p)$ from Eq. \ref{eq:bigG}};
\node (in1) [io, below of=start] {Apply weak object approximation, and disregard higher-order scattering};
\node (pro1) [startstop3, below of=in1] {$
\begin{aligned}
G(\mathbf{K}_f,\mathbf{Q}_p)  = & |A(\mathbf{K}_f)|^2 \delta (\mathbf{Q}_p)\\
& +A(\mathbf{K}_f)A^*(\mathbf{K}_f+\mathbf{Q}_p)\Psi^*_s(-\mathbf{Q}_p)\\
& +A^*(\mathbf{K}_f)A(\mathbf{K}_f-\mathbf{Q}_p)\Psi_s(+\mathbf{Q}_p)
    \end{aligned}
$};
\node (in2) [io, below of=pro1] {For a well-aligned, conventional STEM setup, this now describes 3 discs, of radius $\alpha$, for each $\mathbf{Q}_p$.};
\node (dec1) [decision, below of=in2, yshift=-0.5cm] {${0<\lambda|Q_p|< 2 \alpha}$};
\node (pro2a) [startstop, below of=dec1, yshift=-0.5cm] {Signal transfer possible for this spatial frequency};
\node (pro2b) [startstop2, right of=dec1, xshift=2cm] {No signal transfer for this frequency.};
\node (out1) [io, below of=pro2a] {In the double overlap region (cf. Fig. \ref{fig:overlap} sum the intensity and find the mean phase};
\node (pro3) [startstop, below of=out1] {Assign these values to the $\mathbf{Q}_p$ coordinate, forming a complex 2D array};
\node (pro4) [startstop, below of=pro3] {Perform an inverse Fourier transform, $\Psi(x,y)=\mathcal{F}^{-1}(\mathbf{Q}_p)$, to find the complex specimen transmission function};
\draw [arrow] (start) -- (in1);
\draw [arrow] (in1) -- (pro1);
\draw [arrow] (pro1) -- (in2);
\draw [arrow] (in2) -- (dec1);
\draw [arrow] (dec1) -- (pro2a);
\draw [arrow] (dec1) -- (pro2b);
\draw [arrow] (dec1) -- node[anchor=east] {true} (pro2a);
\draw [arrow] (dec1) -- node[anchor=south] {false} (pro2b);
%\draw [arrow] (pro2b) |- (pro1);
\draw [arrow] (pro2a) -- (out1);
\draw [arrow] (pro3) -- (pro4);
\draw [arrow] (out1) -- (pro3);
\end{tikzpicture}
\caption{Flowchart describing the SSB ptychography analysis pipeline}
\label{fig:ssb}
\end{figure}
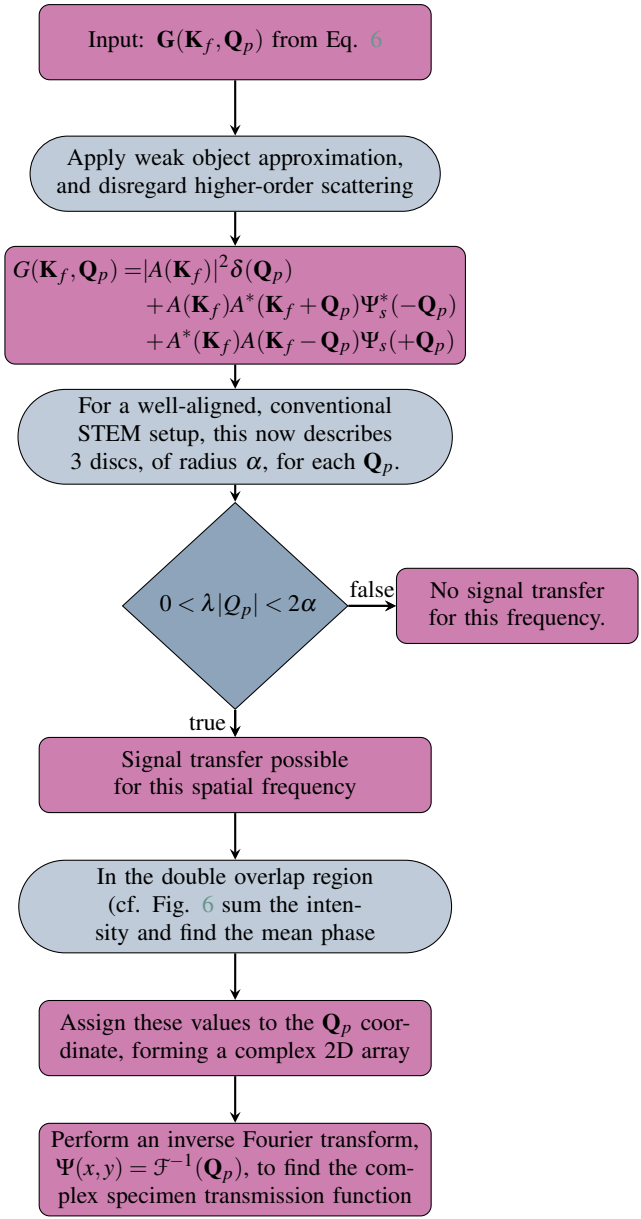

The two most widely used direct methods are single side-band (SSB) ptychography and Wigner distribution deconvolution (WDD) ptychography.
Both the SSB and WDD approaches, or indeed, any other STEM methodology based on the multiplicative object approximation, can be derived from $\mathbf{G}(\mathbf{K}_f,\mathbf{Q}_p)$ as defined in Eq. \ref{eq:bigG}. The algorithm underlying the SSB method was first developed in Ref. \cite{pennycook2015efficient}, and is described in more detail in Ref. \cite{clark2023effect}. This method is described schematically in Fig. \ref{fig:ssb}.

One can interpret this flowchart as follows: The application of the weak phase object approximation (App. \ref{sec:definitions}) enables $\mathbf{G}$ to be simplified to a sum of three-terms. In a conventional STEM these three terms are described by three discs, of radius $\alpha$. One can separate $G$ into four regimes as a function of $Q_p$: where $Q_p=0$ there is no information available as the three discs overlap (referred to as triple-overlap \cite{o2020contrast}) in their entirety. Where $Q_p<\alpha$ the most of the area of these discs are in a triple-overlap regime (as illustrated in Fig. \ref{fig:overlap}(a) )but there are regions where just the central disc and one side disc overlap. This double overlap region is formed by interference between two complex wavefunctions and hence phase information can be transferred. These regions increase in area to a maximum (Fig. \ref{fig:overlap}(b)) and then decrease ($\alpha<Q_p<2\alpha$, until where $Q_p \ge 2\alpha$ there is no region of double overlap (Fig. \ref{fig:overlap}(c)) and hence no information can be transferred. In this way, the phase contrast transfer is optimal where the double overlap region is greatest \cite{seki2018theoretical}.

A specific benefit of the filtering step in the SSB method is efficient noise rejection - in the weak object approximation, signal can only appear in the region of reciprocal space defined by the aperture functions, illustrated in Fig. \ref{fig:overlap}. As such, only forming the image from signals within this region of 4D space removes noise and improves image interpretability at low electron doses \cite{o2020phase}.

The weak object approximation is well-established  to be ill-suited to most samples in high-resolution STEM imaging \cite{cowley1995diffraction}. This may suggest that the SSB algorithm is not of much practical utility in real conditions for samples beyond graphene. This seems not to be the case in practice - recent studies have shown SSB ptychography to perform well for successful atom column determination in relatively thick samples \cite{clark2023effect}, which may be due to the inherent filtering in the selection of the double overlap region step.

A particularly satisfying aspect of the SSB algorithm is that no parameters are needed beyond physical measurements - there is no external judgement of sufficient convergence, or noise filtering parameters.
To the best of our knowledge, this is currently the only ptychographic algorithm with these properties.

\subsubsection{Wigner distribution deconvolution (WDD) ptychography}

That one could obtain the complex specimen transmission function from a 4D-STEM dataset  seems to have been first demonstrated in 1989 (or, as the terminology was then,``all microdiffraction data as a function of probe position") \cite{bates1989sub} - this reference does not use the term ptychography, but does consider deconvolution methods of 4D-STEM data to retrieve an image at higher resolution than is obtainable via conventional imaging methods.

\begin{figure}[t!]
\centering
\begin{tikzpicture}[node distance=1.75cm]
\node (start) [startstop] {Input: $\mathbf{G}(\mathbf{K}_f,\mathbf{Q}_p)$ from Eq. \ref{eq:bigG}};
\node (inA) [io, below of=start] {Apply an inverse Fourier transform with respect to $\mathbf{K}_f$. Note it can be expressed as two Wigner distributions:};
\node (in1) [startstop3, below of=inA, yshift=-0.5cm] {$
\begin{aligned}
     H(\mathbf{R},\mathbf{Q}_p)= &\int a^*(\mathbf{b})a(\mathbf{b}+\mathbf{R}) \, \mathrm{exp}(-2\pi i \mathbf{Q}_p \cdot \mathbf{b}) \, \mathrm{d}\mathbf{b}  \\ \cdot &\int \psi^*(\mathbf{c})\psi(\mathbf{c}+\mathbf{R}) \, \mathrm{exp}(-2\pi i \mathbf{Q}_p \cdot \mathbf{c}) \, \mathrm{d}\mathbf{c}\\
     =& W_a(\mathbf{R},-\mathbf{Q}_p) \cdot W_{\psi}(\mathbf{R},-\mathbf{Q}_p).
 \end{aligned}$};
 \node (inB) [io, below of=in1, yshift=-0.3cm] {Apply a Wiener filter deconvolution \cite{batesbook}, and solve for $W_{\psi}(\mathbf{R},-\mathbf{Q}_p)$};
\node (pro1) [startstop3, below of=inB] {$
\begin{aligned}
W_{\psi}(\mathbf{R},-\mathbf{Q}_p)  = & \frac{W^*_a(\mathbf{R},-\mathbf{Q}_p) \, \, \, H(\mathbf{R},\mathbf{Q}_p)}{|W_a(\mathbf{R},-\mathbf{Q}_p |^2+ \epsilon}
    \end{aligned}
$};
\node (in2) [io, below of=pro1] {Expand $W_{\psi}(\mathbf{R},-\mathbf{Q}_p)$ and Fourier transform:};
\node (dec1) [startstop3, below of=in2] {$D(\mathbf{K}_f, \mathbf{Q}_p)=\mathcal{F}(W_{\psi}(\mathbf{R},-\mathbf{Q}_p))=\Psi(\mathbf{K}_f)\Psi^*(\mathbf{K}_f-\mathbf{Q}_p)$};
\node (pro2a) [startstop, below of=dec1, yshift=-0.5cm] {$\Psi(\mathbf{Q}_p)=\frac{D^*(0, \mathbf{Q}_p)}{\sqrt{D(0,0)}}$};
\node (pro4) [startstop, below of=pro3] {Perform an inverse Fourier transform, $\Psi(x,y)=\mathcal{F}^{-1}(\mathbf{Q}_p)$, to find the complex specimen transmission function};
\draw [arrow] (start) -- (inA);
\draw [arrow] (inA) -- (in1);
\draw [arrow] (in1) -- (inB);
\draw [arrow] (inB) -- (pro1);
\draw [arrow] (pro1) -- (in2);
\draw [arrow] (in2) -- (dec1);
\draw [arrow] (dec1) -- (pro2a);
\draw [arrow] (dec1) -- node[anchor=east] {compress to 2D} (pro2a);
\draw [arrow] (pro2a) -- (pro4);
\end{tikzpicture}
\caption{Flowchart describing a simple implementation of the WDD algorithm}
\label{fig:WDD}
\end{figure}
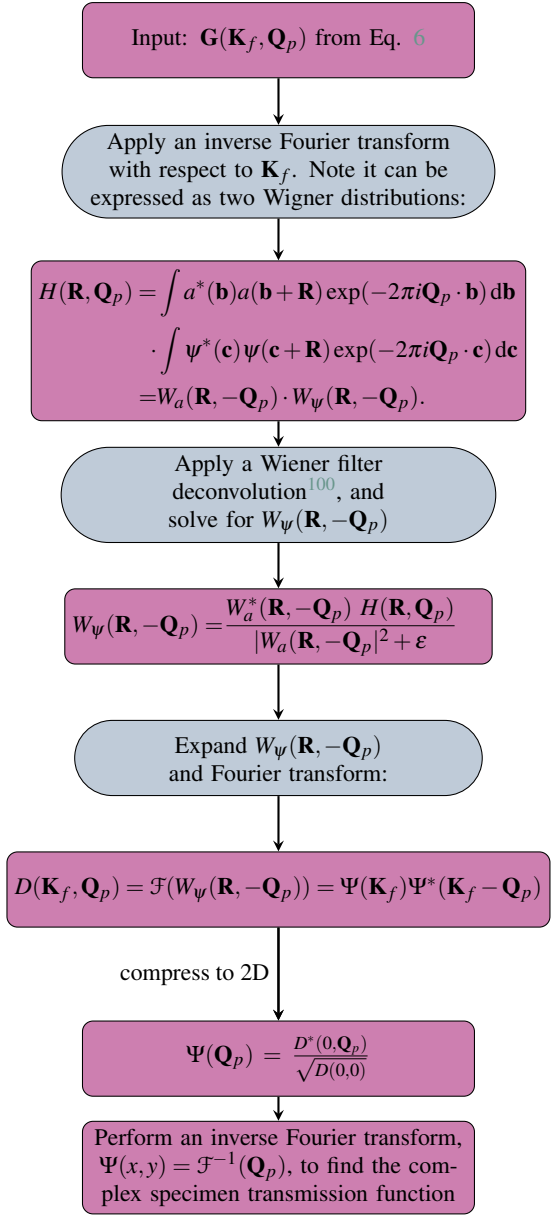

The Wigner distribution deconvolution method (WDD) of ptychography derives from the aforementioned 1989 work, and was first introduced in 1992 \cite{rodenburg1992theory}. The underlying structure of the WDD method is depicted in Fig. \ref{fig:WDD}. The Wiener filter applied in step 4 of Fig. \ref{fig:WDD} requires selection of the parameter $\epsilon$. This is a small value to prevent division-by-zero problems - in Ref. \cite{rodenburg1992theory} approaches to scale this parameter are discussed, while in Ref. \cite{yang2017electron}, a value of 1\% of $|W_a|^2$ is used. The stability and optimal $\epsilon$ choice will vary with dataset and the signal:noise therein. $W_a$ contains only terms characterised by the microscope parameters, and $\mathbf{G}(\mathbf{K}_f, \mathbf{Q}_p)$ is determined from the processed, collected 4D dataset. The deconvolution by Wiener filter is only one deconvolution method that could be applied here - other methods exist and may yet prove more stable and efficient in practice \citep{batesbook}.

The "compress to 2D" step in Fig. \ref{fig:WDD} is only one of the choices available to the scientist - the method presented is merely one simple option to achieve an experimentally useful estimate of $\Psi(\mathbf{Q}_p)$ - achieving the goal at that time, of super-resolution. Alternative methods are suggested in Ref. \cite{rodenburg1992theory}, but have not yet been much explored in experimental settings, and may yet prove a fruitful avenue for further study.

\subsubsection{Other deterministic methods}
While SSB and WDD have both seen fairly wide employment, other methods of ptychography have also been proposed. We note with interest the range of iterative-but-deterministic approaches that have been developed including Refs. \cite{d2014deterministic,martin2008direct}. To the best of our knowledge, these have not seen significant experimental employment, but with advances in detector efficiency and microscope stability, this may yet change.

The direct methods reach their highest possible resolution in a tightly-focused probe geometry. A further benefit of this optical geometry is that one would also be in, or near to,  typical data collection conditions for other STEM methods, including EDX-STEM or HAADF-STEM. Indeed, it is possible to collect these datasets simultaneously to obtain deeper materials insights - discussed further in Sec. \ref{sec:simultaneous}.

\subsection{Iterative methods}
An alternative route to solving the phase problem through a ptychographically-constructed is presented in iterative methods.
An initial guess of the specimen's complex transmission function is made, based on \emph{a priori} knowledge perhaps (such as from a simultaneous bright-field STEM map), or an assumed simple function. From this, given the collected experimental data, and the known range of illumination conditions - one can follow a route similar to the original Gerchberg-Saxton\cite{gerchberg1972practical} (Fig. \ref{fig:GSalg}) using iteration to approach a stable solution.
    \begin{figure}[!ht]
    \centering
    \includegraphics[width=0.8\linewidth]{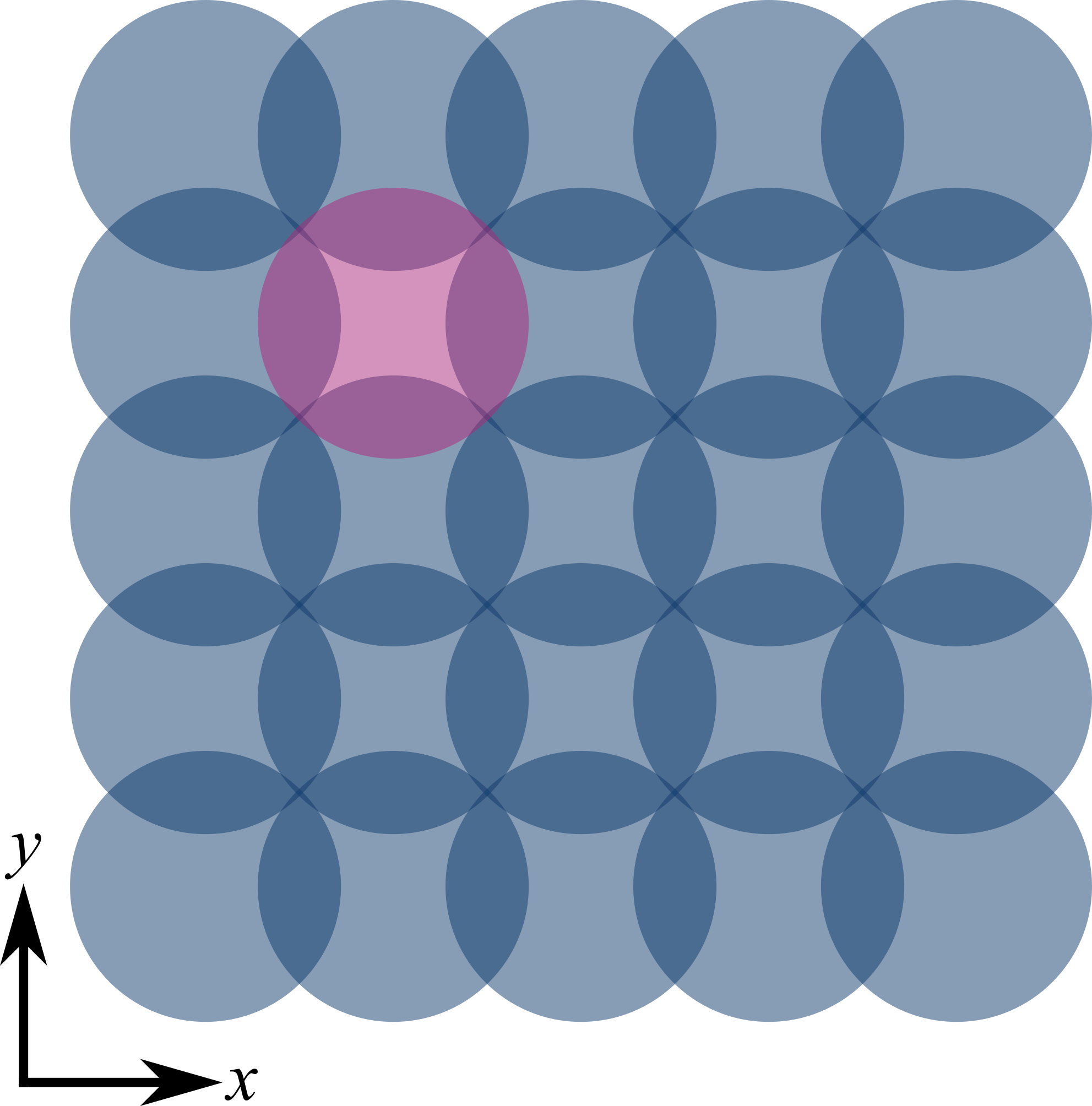}
    \caption{PIE illumination pattern schematic - the set of discs in the sample plane shows the set of probe positions rastered across, in which there are substantial overlaps between probe positions. The highlighted disc shows illumination for one single probe position. We see a significant proportion of the illuminated area is also illuminated by adjacent probe positions.}
    \label{fig:PIE}
\end{figure}

\subsubsection{Ptychographic Iterative Engine (PIE)}\label{sec:PIE}

\begin{figure*}[t!]
 \centering
 \begin{tikzpicture}[node distance=1.75cm]
 \node (start) [startstop] {Setup: Form an initial guess of the complex object transmission function: $O_{guess, n}(x,y)$};
 \node (start1) [startstop, below of=start, yshift=-0.2cm] {Setup: model complex imaging probe, typically a defocused Airy disc, $P(x-X_p, y-Y_p)$};
 \node (start2) [startstop, below of=start1, yshift=-0.5cm] {Model the initial guessed specimen exit wavefunctions for each probe position $p$, as $\psi_{guess, n}(x,y, X_p, Y_p)=O_{guess, n}(x,y)\times P(x-X_p, y-Y_p)$};
 \node (in1) [io, below of=start2, yshift=-0.7cm] {For current $p$, propagate the guessed exit wavefunction to the detector plane by applying a Fourier transform: $\Psi_{guess, n}(k_x, k_y, X_p, Y_p)=\mathcal{F}\left(\psi_{guess, n}(x,y, X_p, Y_p)\right)$};
     \node (in11) [io, below of=in1, yshift=-0.6cm] {Update with experimental modulus: $\Psi_{update, n}(k_x, k_y, X_p, Y_p) = \lvert\Psi_{expt}(k_x, k_y, X_p, Y_p)\rvert \times \mathrm{exp}(i \phi_{guess, n}(k_x, k_y, X_p, Y_p)$};
 %\node (inq) [io, below of=in1] {For probe position $p$};
   \node (in12) [io, below of=in11, yshift=-0.5cm] {Form an updated guess of the exit wavefunction: $\psi_{updated, n}(x,y, X_p, Y_p)=\mathcal{F}^{-1}(\Psi_{updated, n}(k_x, k_y, X_p, Y_p))$};
 \node (pro1) [startstop, below of=in12, yshift=-0.7cm] {Update $O_{guess, n}(x,y)$ (modulated by additional parameters $a$ and $b$ to reduce noise amplification):
     $O_{guess, n+1}(x,y)= O_{guess, n}(x,y) + P_{norm}\times P_{filter} \times T$};
      \node (pro2c) [startstopPALE, right of=pro1, xshift=4cm, yshift=+1.7cm] {$P_{norm}=\frac{|P(x-X_p, y-Y_p)| }{|\mathrm{max}(P(x-X_p, y-Y_p))|}$ localises the update to regions of high probe intensity};
\node (pro2d) [startstopPALE, below of=pro2c, yshift=-0.0cm] {$P_{filter}=\frac{P^*(x-X_p, y-Y_p)}{|P(x-X_p, y-Y_p)|^2+a}$ screens out areas of low signal};
\node (pro2e) [startstopPALE, below of=pro2d, yshift=-0.1cm] {Tuning factor $T = b\times(\psi_{updated, n}(x,y, X_p, Y_p) - \psi_{guess, n}(x,y, X_p, Y_p))$ improves stability in noisy datasets};
 \node (decA) [decision, below of=pro1, yshift=-0.6cm, aspect=2] {$p=p_{end}$?};
 \node (decB) [decision, below of=decA, yshift=-0.8cm, aspect=2] { $O_{guess, n}(x,y) \approx O_{guess, n-1}(x,y)$?};
\node (fin) [startstop, below of=decB, yshift=-0.8cm] {Iterations complete};
 \draw [arrow] (start) -- (start1);
 \draw [arrow] (start1) -- (start2);
 \draw [arrow] (start2) -- (in1);
 \draw [arrow] (in1) -- (in11);
 \draw [arrow] (in11) -- (in12);
  \draw [arrow] (in12) -- (pro1);
\draw [-] (pro1) -- (pro2d);
\draw [-] (pro2d) -- (pro2c);
\draw [-] (pro2d) -- (pro2e);
\draw [arrow] (pro1) -- (decA); 
\draw [arrow] (decA) -- node[anchor=east] {Yes} (decB);  
\draw [arrow] (decB) -- node[anchor=east] {Yes}(fin);  
 \draw[arrow] (decA.west) -- ++(-1.8,0) -|  ++(0,+9.4) node[near end ,anchor=south, rotate=90] {No, $p=p+1$} -- (in1.west);
  \draw[arrow] (decB.west) -- ++(-2.2,0) -|  ++(0,+14.4) node[near end ,anchor=south, rotate=90] {No, $p=1$, $n=n+1$} -- (start2.west); 
 \end{tikzpicture}
 \caption{Flowchart describing the PIE algorithm as a ptychographic approach to solving for the complex specimen transmission function, or exit wavefunction: $O(x,y)$.}
 \label{fig:PIEmethod}
 \end{figure*}
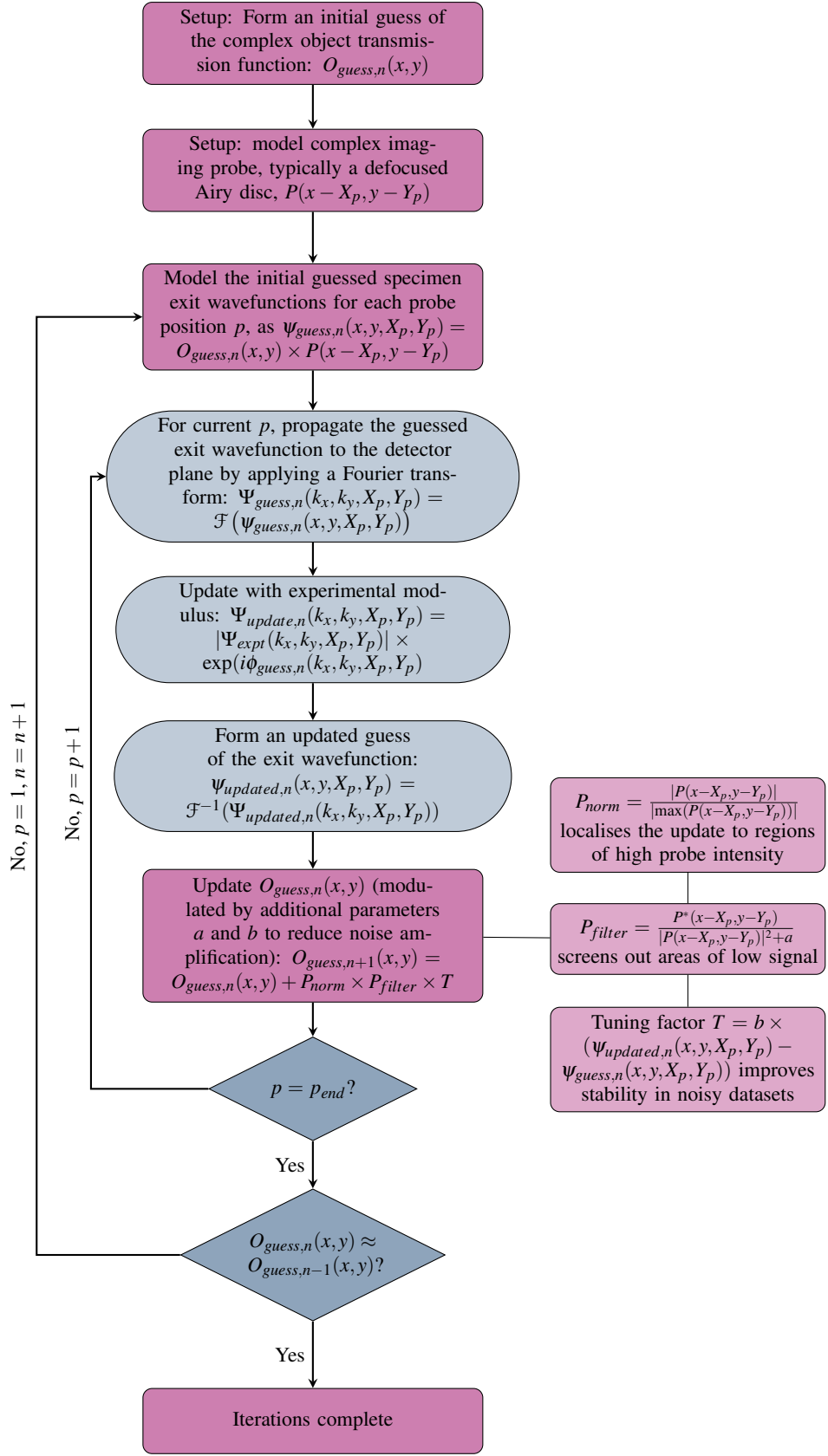

The first steps towards an algorithm of this sort were made by Faulkner and Rodenburg \cite{faulkner2004movable}, with a moveable aperture approach to ensure localised regions of illumination with a defined edge. In this approach, one can see similarities to earlier work in coherent diffractive imaging in which having a finite-bound to the update area is key to enabling robust convergence \cite{hawkes_spence_2007}. It was not long thereafter, that the same authors realised their algorithm could also work with a more conventional (STEM-like) illumination profile \cite{rodenburg2004phase}, with a softer-edged update area. This method, now referred to as  PIE - notionally defined as the Ptychographic Iterative Engine, but also a reference to an early description of ptychographic methods as "pie in the sky" \cite{rodenburg2019ptychography}\footnote{This is an English idiom suggesting scepticism of the idea \cite{piesky}}. If one collects a 4D-STEM dataset from overlapping probe positions, as illustrated in Fig. \ref{fig:PIE},  and one knows or can determine the complex probe wavefunction $P(x-X_p, y-Y_p)$, which only shifts laterally  during the scan, across a grid of scan positions described by the set $(X_p, Y_p)$, the PIE algorithm proceeds as illustrated in Fig. \ref{fig:PIEmethod}.

PIE developed further, and was subsequently demonstrated by Rodenburg \emph{et al.}\cite{rodenburg2007transmission}. The multiple-illumination setup of PIE provides data-diversity in that areas are repeatedly illuminated as the step size between probe positions is smaller than the diameter of the illumination impinging upon the specimen. For manageable data sizes, this is usually achieved through defocussing the probe such that the probe diameter is rather large and so fewer probe positions are needed to span the region of interest. This real-space requirement of the illumination condition is illustrated schematically in Fig. \ref{fig:PIE}). These overlaps support algorithm convergence and helps to reduce the occurrence of the algorithm getting stuck in a local minima of solution space. We note however, that the definition of "sufficient" overlap in the probe illumination profiles is non-trivial \cite{rodenburg2019ptychography} and is a parameter worthy of optimisation for any given imaging setup. The overarching PIE method is illustrated in the flowchart of Fig. \ref{fig:PIEmethod}, while in the following discussion, we give more guidance on each of the steps within the PIE algorithm.

The initial guess of the complex object transmission function $O_{guess, n=1}(x,y)$ could theoretically take any form (i.e., planar amplitude and phase, random noise, or a close approximation to the expected object transmission function). In practice, a poor choice of initial object transmission function can lead to the algorithm output stagnating in a local minima of solution space \cite{rodenburg2019ptychography}.

The input model for the probe is typically simple to define in the STEM, as the Fourier transform of the condenser aperture, with an applied aberration function (i.e. non-zero defocus): 
${P(x,y)=\mathcal{F}(A(k_r < \alpha) \chi(k_r, k_{\phi}))}$. This is then shifted across each of the $(X_p, Y_p)$ probe positions in the raster scan (as illustrated in Fig. \ref{fig:PIE}). While a focused probe without aberration forms the conventional Airy disc profile \cite{clark2018probing}, a defocussed probe tends towards a top-hat intensity profile.

The value of defocus can be freely chosen - balancing step size with probe size for datasize management, but also considering the fineness of details in the expected diffraction pattern with the number and relative size of pixels in the available detector. For examples of chosen values, in Ref. \cite{zhou2020low}, $df=-13 \mu$m was used to achieve high-resolution imaging of virus and virus-like particles, while in Ref. \cite{humphry2012ptychographic}, a defocus of approximately $1.7\mu$m was applied - acheiving atomic-resolution maps at 30keV. We note however, that in-focus applications can also work and enable simultaneous acquisition of analytical or conventional imaging methods \cite{allen2023super, clark2023effect}. A detailed discussion of parameter choice can be found in Ref. \cite{blackburn2021practical}. In general, a suitable defocus value will vary with coherence length, sample beam-damage behaviours, and  optical stability of the microscope along with detector characteristics (MTF, number of pixels, response speed) as a sampling limit \cite{rodenburg1992theory}.

The experimental modulus is the square root of the recorded intensity pattern from probe position, $p$. In the case of a detector that is not scaled to give output in terms of incident electrons, this intensity (and all related functions) can be normalised to an arbitrary intensity (e.g. normalise the array to a total intensity of 1 in arbitrary units). The $a$ parameter, appearing in $P_{filter}$ is a small, positive real value introduced to avoid divide by zero errors.

In their derivation, Rodenburg and Faulkner \cite{rodenburg2004phase} astutely note that it is the introduction of the tunable parameters $a$ and $b$ which differentiate this method from earlier, similar phase retrieval methods. It is also this additional flexibility which enables the robustness of this algorithm to variations found in experimental datasets.

Since the introduction of PIE, many more iterative algorithms have been developed to improve algorithmic reliability under various experimental limitations. These algorithms can be broadly classified as PIE-variations (Sec. \ref{sec:ePIEetc}) and non-PIE iterative tools (Sec. \ref{sec:notPIE}). With these tools available, experimental ptychographic algorithms can now account for a range of challenges such as varying aberrations, dose limitations, and uncertainties in experimental measurements. These enable practical applications of ptychography in a range of interesting and challenging experiments - some of these are highlighted below in Section \ref{sec:applications}.

\subsubsection{Alternative flavours of PIE}\label{sec:ePIEetc}

PIE, while a powerful algorithm, suffers from some key limitations which can be present in real STEM experiments.

In the STEM, we do not always have an accurate model for our probe as it impinges on the sample (due to e.g. aberrations, aperture charging, sample-height variation induced defocus). In 2008, Thibault and colleagues \cite{thibault2008high} demonstrated an extension to the PIE method in their work in STXM (scanning transmission X-ray microscopy - an imaging method which can be considered approximately as the X-ray equivalent to a STEM setup) in which an alternating projections approach is taken to the problem, enabling iterative updates of the modelled transmission function, and the modelled probe, showing improved robustness to local minima. This work was swiftly followed in 2009 by Maiden and Rodenburg \cite{maiden2009improved}. In Ref. \cite{maiden2009improved}, the authors developed the extended PIE (ePIE) algorithm to help overcome the same limitations of the original PIE - adding a further update step in a serial approach, allowing updates of the modelled probe as well as the sample - this algorithm has found wide application in electron microscopy, while the approach of Ref. \cite{thibault2008high} finds wide application in the STXM community (at the time of writing, both of these articles have >1,300 citations per Google Scholar). 

While in Fig. \ref{fig:PIEmethod}, we see the update step: $O_{g, n+1}(x,y)= O_{g, n}(x,y) + P_{norm}\times P_{filter} \times T$, which in long form is expressed below, Maiden and Rodenburg introduced an additional step, performed subsequently to each sample update, one probe position at a time $P_{g, n+1}(x,y)$:
\begin{widetext}
\begin{equation}
    O_{g, n+1}(x,y)=  O_{g, n}(x,y) + \frac{|P(x-X_p, y-Y_p)| }{|\mathrm{max}(P(x-X_p, y-Y_p))|}  \frac{P^*(x-X_p,  y-Y_p)}{|P(x-X_p, y-Y_p)|^2+a} \times b\times(\psi_{updated, n}(x,y, X_p, Y_p) - \psi_{g, n}(x,y, X_p, Y_p))
\end{equation}
\begin{equation}
    P_{g, n+1}(x,y)=  P_{g, n}(x,y) + \frac{|O(x+X_p, y+Y_p)| }{|\mathrm{max}(O(x+X_p, y+Y_p))|}  \frac{O^*(x+X_p,  y+Y_p)}{|O(x+X_p, y+Y_p)|^2+c} \times d\times(\psi_{updated, n}(x,y, X_p, Y_p) - \psi_{g, n}(x,y, X_p, Y_p)).
\end{equation}
\end{widetext}
In this probe update, we see an elegant mirroring of the object update step - an approach which has proven remarkably effective in the huge range of experimental conditions to which it has been applied subsequently. In particular, in the latter sections of Ref. \cite{maiden2009improved}, the authors investigate convergence under a range of conditions relevant to experimental electron microscopy and find ePIE to be rather robust. In Ref. \cite{maiden2009improved} we also see a re-emphasis of the engineering-approach to ptychography - combining a few cycles of one algorithm, before a few of another in order to overcome local minima and support convergence towards a global solution (as suggested earlier by Fienup \cite{fienup1978reconstruction} - a pragmatic approach that is perhaps underappreciated outwith the small community of ptychographers. The ePIE method was applied very successfully by Jiang \emph{et al.} \cite{jiang2018electron}.

These methods however still assume accurate knowledge of the probe position relative to the sample across the raster scan. As such, the presence of sample drift during a scan can limit the success of ptychographic reconstruction in such cases. The noise levels in the recorded dataset can be a limiting factor, as to how accurate the initial model of the probe and object functions must be for a successful convergence to a correct solution.

\subsubsection{Solution stability \emph{(or why PIEs should be round)}}\label{sec:ConvexSets}

\begin{figure}[!ht]
  \centering
  \subfloat[][]{\includegraphics[width=.4\linewidth]{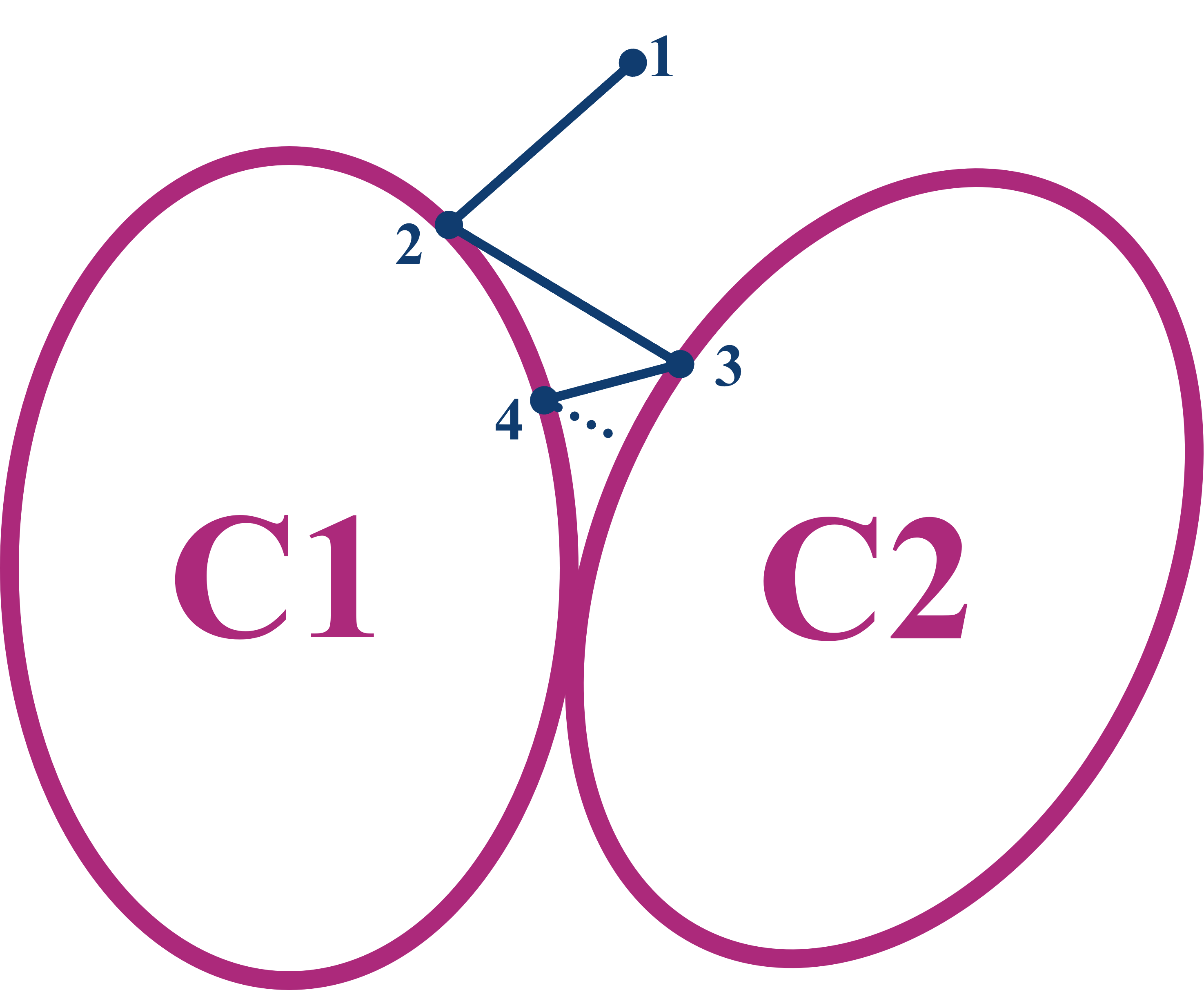}}\quad
  \subfloat[][]{\includegraphics[width=.4\linewidth]{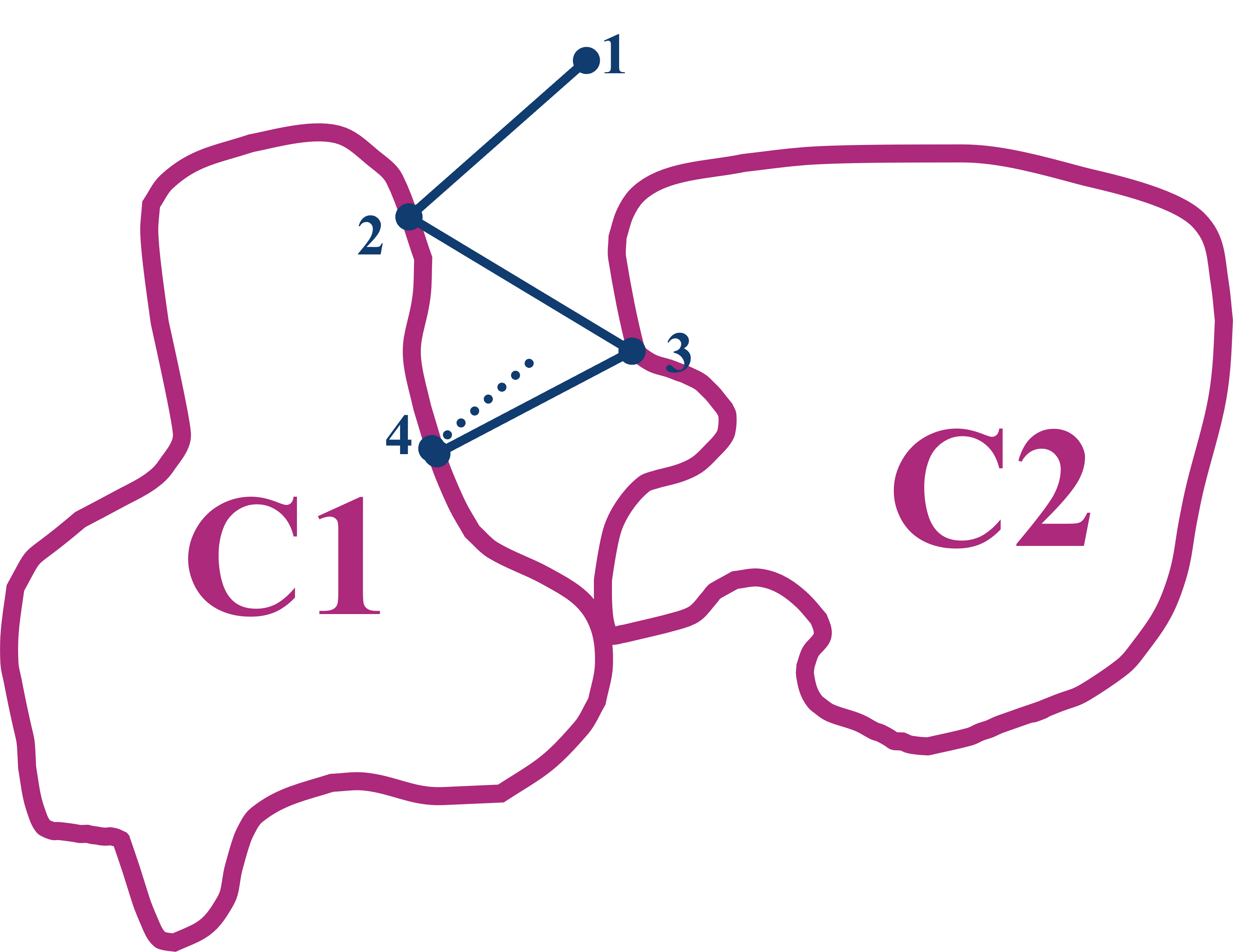}}\\
  \subfloat[][]{\includegraphics[width=.4\linewidth]{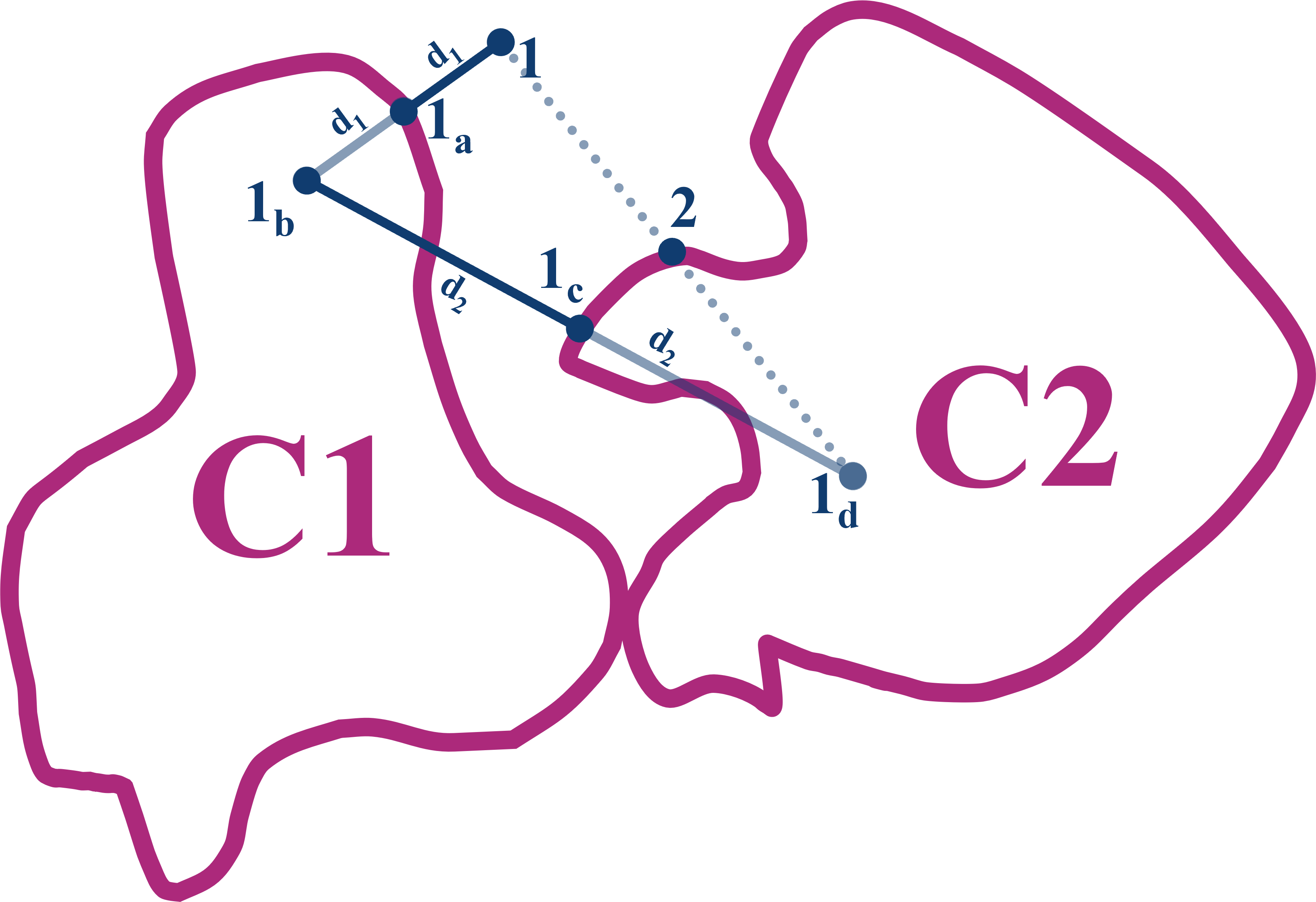}}
     \caption{(a) Given noiseless convex sets, our iterative solvers can progress smoothly to the true solution (the overlap where both constraint sets are satisfied). (b) In all real models, the constraint spaces will have some degree of non-convexity, in which case the found-solution may rapidly diverge from the vicinity of the true answer. (c) Using a projective Difference-Map type approach, the solver robustness to non-convexity is improved, through this smoothing process. C1 indicates the region in which constraint set 1 is satisfied, while C2 indicated the region in which constraint set 2 is satisfied.}
     \label{fig:convexsets}
  \end{figure}

The experimental success of PIE relies on each iteration bringing the solution closer to the true answer. Successful iteration is not guaranteed in reality - noise and other experimental errors force the parameter space to be non-convex \cite{cover1999elements}, in which case one finds the found-solution may diverge from the true answer. Repeating the iterative process with alternative initial guesses may provide a route forwards in some cases. The challenge posed by non-convex sets is illustrated schematically in Fig. \ref{fig:convexsets} (following Ref. \cite{rodenburg2019ptychography})

\subsubsection{Other iterative approaches}\label{sec:notPIE}

There are other approaches now available to solving the phase problem using ptychographic / deconvolution-based tools. Here we detail and discuss some others of the key algorithms available to, and potentially useful for, the electron microscopy community. We aim here for a broad coverage of the conceptual variants available, with a bias towards those currently widely employed for electron microscopy studies.

\begin{itemize}
    \item Difference map
    
The difference map approach, introduced by Elser in 2003 \cite{elser2003phase} enables a much broader understanding of iterative solution approaches as applicable to phase-retrieval \cite{elser2007searching}. Indeed, Elser recognises the hybrid input-output method as one simple approach among a multidimensional array of potential routes to phase retrieval. The difference map method can be used for conventional phase retrieval (i.e. without a full 4D dataset) amongst many possible applications, but can flexibly be adjusted to our requirements. This approach was used to great effect by Thibault \emph{et al.} in 2008 \cite{thibault2008high}.

Inspection of Fig. \ref{fig:convexsets}, reveals that each step (from Point \raisebox{.5pt}{\textcircled{\raisebox{-.9pt} {1}}} to Point \raisebox{.5pt}{\textcircled{\raisebox{-.9pt} {2}}} in each of the sub-figures) in the difference map algorithm is a little more involved than the previous methods, fewer iterative steps are overall are expected to be needed.  The difference map process follows Fig. \ref{fig:convexsets}c as follows: from an initial guess \raisebox{.5pt}{\textcircled{\raisebox{-.9pt} {1}}}, apply the changes required to the data to apply the first constraint \raisebox{.5pt}{\textcircled{\raisebox{-.9pt} {1a}}}. Double this effect, reflecting further into the solution space of constraint set one \raisebox{.5pt}{\textcircled{\raisebox{-.9pt} {1b}}}. From here, apply the second constraint, reaching point \raisebox{.5pt}{\textcircled{\raisebox{-.9pt} {1c}}}. Double this, to reach point \raisebox{.5pt}{\textcircled{\raisebox{-.9pt} {1d}}}. Find the midpoint between initial point \raisebox{.5pt}{\textcircled{\raisebox{-.9pt} {1}}} and current point \raisebox{.5pt}{\textcircled{\raisebox{-.9pt} {1d}}}. This is now point \raisebox{.5pt}{\textcircled{\raisebox{-.9pt} {2}}}, the starting point of our next iterative sequence.

    \item Relaxed averaged alternating reflections (RAAR)
RAAR, a similar optimisation approach to that of the Difference Map algorithm, was first introduced as a ptychographic imaging technique in 2004 \cite{luke2004relaxed}, with further discussion and insight brought by the analyses of Yang \emph{et al.} \cite{yang2011iterative}, and brought to the wider awareness of the electron ptychography community by Rodenburg and Maiden \cite{rodenburg2019ptychography}.

\item Maximum likelihood (ML)
The maximum likelihood approach to solving ptychographic problems was first presented by Thibault and Guizar-Sicairos in 2012 \cite{thibault2012maximum}, with further computational accelerations and flexibility presented in 2018 \cite{odstrvcil2018iterative}.
In this approach, the task is rather decoupled from the optical model, and instead structured as a pure optimisation problem in the different data-spaces.

The flexibility enabled by the ML approach was used to great effect in Refs. \cite{chen2021electron, chen2020mixed} to overcome experimental uncertainties and increase the experimentally attainable resolution.
\end{itemize}

Other methods%, including  generalised gradient descent and Conjugate gradient least squares, and undoubted many other
exist amongst the broad literature - here we cannot possibly cover a full survey of all methods and so have instead selected key approaches to highlight.

\subsection{Mixed-state ptychography}
A particular benefit of the ML approach however, is that this formalism also allows for separation of the mixed-coherence states present in any real imperfectly coherent probe - and as such - a route to attaining higher resolution measurements of our samples. The quantum tomographic state view required for this was first realised by Thibault and Menzel in 2013 \cite{thibault2013reconstructing}, following the 2012 ML work \cite{thibault2012maximum} - and indeed is key to the underlying improved resolution of subsquent applications of this method in electron microscopy.

\subsection{Multislice STEM ptychography\label{sec:multislice}}
To this point in the review, we have solely considered samples which fulfil the multiplicative approximation and thus a single 2D array of data is the output sought. However, this model is not suitable for all samples of interest - and ptychography provides a route to overcome this, moving beyond this simple model.

First demonstrated by Maiden \emph{et al.} in 2012, \cite{maiden2012ptychographic}, the multislice ptychographic model works as a blend of the ePIE algorithm, with the multislice model of electron wave propagation in the electron microscope \cite{kirkland2020advanced}.

Subsequent work employing this approach includes Ref. \cite{gao2017electron} by Gao \emph{et al.}, and then atomic-resolution in three-dimensions demonstrated by Chen \emph{et al.} \cite{chen2021electron}. Wider application of this method is limited in some cases by the current high-dose and high-stability requirements, but this remains a very promising avenue of future developments.

\subsection{Ptychography with plane wave illumination: Fourier Ptychography \label{sec:FourierPtych}}

Some of the early work that would nowadays be described as Fourier ptychography was based around TEM-type illuminatation \cite{faulkner2004movable, haigh2007super, haigh2009atomic}, with more recent electron ptychography showing a STEM geometry bias, partly due to the more reliable experimental stability of a STEM-scan when compared to a set of tilt-conditions in TEM, and partly due to the potential for simultaneous imaging modalities enabled by the localised illumination.

Fourier ptychography is widely used in allied imaging techniques \cite{zheng2021concept, horstmeyer2016diffraction, wakonig2019x} but not yet in the transmission electron microscope. Some recent work however has made approaches in that direction - with both Fourier precession work \cite{lorenzen2024imaging} and a tilt-correction technique \cite{yu2024dose}.

\subsection{Sampling, sensitivity and precision}
In gathering the dataset for ptychographic analyses, we have a multidimensional parameter space available to us to maximise the sensitivity and precision of the data through tuning our sampling in both real and reciprocal spaces. This provides opportunities to ensure that the information we seek about our sample is transferred to the collected signal with a sufficiently robust signal-to-noise ratio to achieve the resolution sought.

While in many experiments the total dose is limited by sample beam-sensitivity \cite{ilett2020analysis}, or by microscope drift and stability considerations. The certainty of any measurement increases with dose (as per Poisson statistics). In modern counting detectors, other noise sources can be neglected - and so will not be considered further here.

In the non-ptychographic but related imaging method of differential phase contrast STEM \cite{clark2018probing}, the Poisson noise in the data propagates through to the final image in an analytically tractable manner \cite{seki2018theoretical}. In the direct methods of ptychography, the propagation of information and noise from recorded data to final image is spatial-frequency dependant \cite{o2020phase} and has not yet been fully fomulated in an analytical model - however we know in the case of SSB, that the strongest signal is obtained for the image spatial frequncy equal to the semi-convergence angle as this leads to the largest area of double overlap.

Then further still we can consider the magnification of the dataset across the pixels of the detector plane - experimentally tunable with the camera length, and limited by the number of detector pixels from the theoretical lowest limit of 3 \cite{brown2016structure} to those with over $1000\times1000$ pixels. In the focused probe geometry where the algorithms require overlap in reciprocal ($q_x, q_y$) space a sampling giving many pixels across the bright-field disc is not particularly impactful above a lower limit of around 16 pixels \cite{yang2015efficient}, due to the dense real-space sampling due to small probe step-size, while in the defocused probe geometry (used for ePIE type methods) the overlap is needed in the real-space of probe illumination on the sample, and the ability to detect small features in the disc (long camera length for large magnification of the bright field disc) will directly enhance the resolution achievable in the final image.

\begin{figure}
     \centering
     \begin{subfigure}[b]{0.45\linewidth}
         \centering
         \includegraphics[height=2cm]{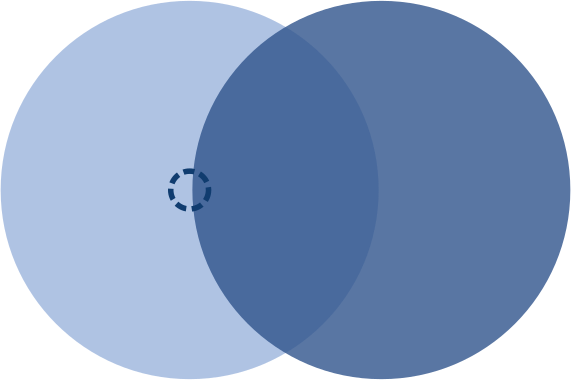}
         \caption{Bright field STEM case}
         \label{fig:resBF}
     \end{subfigure}
     \hfill
     \begin{subfigure}[b]{0.45\linewidth}
         \centering
         \includegraphics[height=2cm]{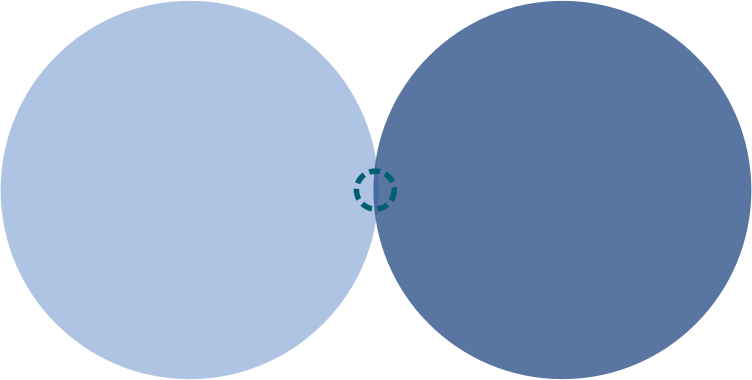}
         \caption{Direct ptychography case}
         \label{fig:resSSB}
     \end{subfigure}
     \caption{Schematics of the maximum resolution attainable in (a) BF-STEM and (b) (non-superresolved) focused probe ptychography. In each case the lighter disc represents the unscattered beam, the darker disc represents the scattered beam at the maximum detectable scattering angle, and the small dashed ring shows the position of the detector required for that scattering angle (and thus, image frequency) to be detected.}
\end{figure}

Resolution in direct methods of phase retrieval (i.e. CoM, SSB, WDD) the maximum spatial frequency that can be represented in an image is defined by a contrast transfer limit of $2 \alpha$, as illustrated schematically in Fig. \ref{fig:resSSB}, in contrast to conventional bright-field STEM, limited to a $1\alpha$ limit (Fig. \ref{fig:resBF}) \cite{hawkes2019springer}. The locality of sample information is restricted to the experimental probe step-size - the resultant output dataset has only the same number of pixels as the number of probe positions in the scan (we note however, that such an image can be trivially up-sampled in a perfectly periodic system).

The image resolution in an iterative approach however, is rather different, and rather difficult to define with simple equations. As a first step, one can decouple (significantly) the size of the image array (in pixels) from the number of probe positions.

In conventional bright-field imaging in STEM, the maximum scattering angle that can be represented in an image is limited by the semi-angle subtended by the probe-forming aperture. That is:
\begin{equation}
    k_{max}=\frac{\alpha}{\lambda}
\end{equation}
This is illustrated in Fig. \ref{fig:resBF}, in that to contribute to the BF contrast, the outermost edge of the scattered beam must still land on the small central detector which lies axially under the position of the unscattered beam - the resolution is scaled to the radius of the beam \cite{pennycook2011scanning}\footnote{The reader may note that there is a typographical error in Ref. \cite{pennycook2011scanning}, p94, and the equation cited is inverted.}.
 Alternatively, the equation for the maximum scattering angle $k_{max}$ can be inverted and interpreted instead as a minimum resolvable distance for in-plane imaging, $d_{min}$. (Axial resolution will be discussed in the section on optical sectioning and multislice methods \ref{sec:algorithms}).
 
 For an aberration-corrected STEM, one might typically have $E=200$keV (i.e. $\lambda=2.5$\~pm), paired with a semi-convergence angle of $\alpha=21$\~mrad, giving a resolution limit of $d_{min}=1.2$ \AA\ for BF-STEM imaging in this example.

Ptychographic algorithms provide us with routes to overcome this conventional resolution limit - providing microscope illumination and sample stability are sufficient (see discussion above).

The resolution limit achieved from a dataset depends on the algorithm through which the recorded 4D-STEM data is processed. In the simplest application of the direct methods (namely SSB and WDD ptychography), one doubles the achievable resolution limit (i.e., $d_{min}=0.5\frac{\lambda}{\alpha}$ in these cases), without the use of any further super-resolution tools.

This twofold gain in resolution is geometrically justified in Fig. \ref{fig:resSSB}: the highest scattering angle that can contribute to the non-superresolution version of direct focused-probe ptychography is proportional to the diameter of the bright-field disc - where the outer edge of the bright-field disc can interfere with the innermost edge of the scattered beam. That is in these methods, we have $k_{max}=\frac{2 \alpha}{\lambda}$ (recalling that $\alpha$ is the beam semi-convergence angle, and thus the radius of the discs illustrated in Fig. \ref{fig:resSSB}). For the typical values described above ($E=200$keV, $\lambda=2.5$pm and $\alpha=21$mrad), for a straightforward ptychographic analysis one could achieve a minimum resolvable distance of $d_{min}=0.6$ \AA\ . In this way, ptychographic analyses allow high-resolution microscopes to achieve atomic-resolution imaging for challenging crystallographic orientations - and for atomic-resolution imaging to become feasible for less well-corrected instruments (methods of computational aberration-correction using related analyses are discussed in Sec. \ref{sec:applications}).

\section{\label{sec:applications}Applications of Electron Ptychography}
While much of the interest in electron ptychography in recent years has been in the developments of the methodology, there are now also a range of publications in which applying ptychographic methods have enabled new materials insights which were previously unachievable. Here, we highlight a few of these. % and note that further discussion of applications of STEM-ptychography can be found in Ref. \cite{SanchezNanoLett}.

\subsection{Low dose ptychography for imaging beam-sensitive materials, and samples with low- atomic number elements}
One particular application family of interest is to beam-sensitive materials. While it seems that TEM may ultimately have the overarching dose-efficiency over STEM techniques \cite{dwyer2023quantum}, for many questions of interest, STEM is the more powerful, adaptable approach. Within the family of STEM methods, ptychography has been demonstrated, both mathematically \cite{o2020contrast} and experimentally \cite{pennycook2019high, lozano2018low, zhou2020low, chen2020mixed} to be highly dose efficient and a practical method to image a variety of materials - across the physical, materials and biological sciences.
As such, ptychography is an increasingly important toolkit for these beam-sensitive materials and holds a great deal of promise for increasing the resolution at which we can image beam-sensitive materials.

\subsection{Lorentz ptychography}
Lorentz imaging in the (S)TEM is the manner of imaging the magnetic structure of materials in a field-free environment. In conventional (S)TEM , as aligned for high-resolution, the electromagnetic lenses used to focus the beam also pass very high magnetic fields of around 2T through the sample - typically erasing the magnetic structure of interest.
The established route around this, is to operate without the primary imaging lenses, accepting a lower resolution instead, nearer to 10nm rather than the usual 0.1nm of STEM - and detect the effects of the sample magnetism on the phase of the sample transmission function by differential phase contrast methods \cite{rose1976nonstandard}.

However, with the superresolution approaches of ptychography, some of this resolution can now be regained \cite{chen2022lorentz, you2023lorentz}.

\subsection{Acquiring complementary data simultaneously with ptychography} \label{sec:simultaneous}

While 4D-STEM approaches to materials science imaging allow us to collect the intensity of all the scattered electrons passed through a sample, and thus making a rather efficient use of the available data, there are further approaches too allowing a more holistic utilisation of the available information. Depending on the materials science question at hand, this may prove a useful route towards the required answer.

Recent examples of this include ptychography paired with tomographic imaging \cite{pelz2023solving} - and indeed, this combination of two large-data imaging approaches leads to significant data sizes requiring careful data management approaches. This pairing of methods however enables imaging of light and heavy elements (low and high atomic numbers) simultaneously via the ptychographic algorithms and an understanding of their 3D arrangement (via the tomographic algorithms).

Ptychographic methods can also be combined with a conventional high-angle annular dark-field (HAADF)-STEM detector, enabling a dense sampling of the bright-field disc and high scattering-angle data collection without a resultant huge dataset \cite{yang2017electron}.

Further combinations of techniques, such as simultaneous EELS and ptychography have been demonstrated theoretically \cite{song2018hollow} although there is yet to be a demonstration of a pixellated detector with the central portion removed to  enable this approach experimentally.

\subsection{Aberration correction}

Saved for the final of the recent demonstrations of experimental electron ptychography is a discussion of aberration-correction, post-acquisition.

From the earliest papers on electron ptychography, the goal to disentangle and 'unfold' microscopy effects from sample information was in mind \cite{Hoppe_1969_I, rodenburg1992theory} - a lofty and worthwhile goal.

Early experimental work by PDN made some practical progress towards this with the sensitivity to aberration in the data and some filtering possible \cite{nellist1994beyond}.

Twenty years later, subsequent work demonstrated this process on the new generation of electron detectors, allowing an iterative procedure to be developed and demonstrated \cite{yang2016simultaneous}.

Then the final hurdle was recently cleared by Nguyen \emph{et al.} with their publication "Achieving sub-0.5-angstrom–resolution ptychography in an uncorrected electron microscope" in which ptychographic data from a non-corrected electron microscope was able to match the resolution achieved with data from a (significantly more expensive) aberration corrected microscope, achieving sub-0.5\AA resolution \cite{nguyen2024achieving}. At this stage, it is too early to confirm how widely these methods will be taken up and found to be practiable across a wide range of experimental setups - but the results are highly-encouraging for the future of electron ptychography.

\section{\label{sec:Conc}Summary and Conclusions}

In this review article, we have described the overarching framework of ptychography as employed in the (scanning) transmission electron microscope, and worked through the primary methods in detail. We have given guidance as to their limitations and the relative capabilities of different genres of ptychography algorithm, and then followed with a range of applications of ptychography to important materials science research questions of the day. Ptychography has developed hugely from the initial inklings of Prof. Sir. Peter B. Hirsch, to its developments by Hoppe and colleagues, and Rodenburg and since the turn of the millennium has grown to a great smorgasbord of techniques providing an appropriate toolkit for a great many scientific research questions.

The development of ptychography is still on-going with key questions, as of yet, unanswered: There is no fundamental proof that the iterative methods will converge to a correct answer - though it works rather well in practice. The stability and applicability of the multislice methods hold much promise but we eagerly more practical implementations. Which algorithm is most dose-efficient, and if ptychography can overcome the quantum imaging limit as laid out recently \cite{dwyer2023quantum}. Whether the phase precision of ptychography can out-perform that of holography. Perhaps in the coming decade one can hope for answers to some of these.

\begin{acknowledgments}
%The authors are grateful to Angela Kirby for guidance on international phonetic alphabet transcription.

LC acknowledges funding from a Royal Society University Research Fellowship (URF\textbackslash R1\textbackslash 221270) and additional Royal Society funding (RF\textbackslash ERE\textbackslash 221035).

PDN acknowledges funding from the Henry Royce Institute for Advanced Materials (EP/R00661X/1, EP/S019367/1, EP/R010145/1) and grant number EP/M010708/1.  Financial support was also received from the EU H2020 Grant Number 823717 ESTEEM3. 
\end{acknowledgments}

\section*{Data Availability Statement}
Data sharing not applicable – no new data generated

\appendix \section{Definitions}\label{sec:definitions}

  \textbf{The phase problem} describes the phenomenon common to multiple imaging sciences that one can only directly measure the intensity of a complex wave, while one seeks the amplitude \emph{and phase} of the wave to fully understand the system under study \cite{taylor2003phase}.
 The phase problem is a catchphrase, describing that for many imaging techniques, it is only possible to collect the amplitude of a complex wave and not the full amplitude and phase which fully determine the wavefunction.

\textbf{The multiplicative object approximation} describes a sample as a single transmission function, $T(x,y)$ which can modify both the amplitude and phase of the transmitted function 
\begin{equation}
    T(x,y)=A(x,y)\mathrm{exp}(i \phi (x,y)
\end{equation}

\textbf{A phase object} describes a sample which we can make the multiplicative object approximation, and then further assume that $A(x,y)=1$. 

The \textbf{weak phase object approximation} goes a step further, to then assume that $\phi(x,y)$ is "small", and as such the sample only makes a small (usually linear) change on the electron wavefunction as it passes through the sample.

Similarly, a \textbf{thin object} is when one can assume the probe does not change profile within the sample substantially - the converse being the case is a \textbf{thick object}. At higher convergence angles, a focussed electron probe has a shorter depth of focus \cite{hecht2012optics}, and thus more samples are thick.

With these terms, we hope that the reader can find their way through the key aspects of the methods described in this article and references therein.

\section{Available open access codes for ptychography\label{sec:codes}}

Recent years have also seen an uptake of open-access philosophies amongst the computational imaging community, with more and more codes being made openly available.  As a guide to the interested reader, here we provide details of a number of packages suitable for electron ptychography, along with brief descriptions of their features. This is not, and can not be, a measure of their accuracy or validity - but those mentioned have been employed in publications cited in this review.

\begin{itemize}
    \item PtychoSTEM (Matlab) \url{https://gitlab.com/ptychoSTEM/}
    \item PyPtychoSTEM (Python) \url{https://gitlab.com/pyptychostem/pyptychostem}
    \item Ptypy python \url{https://ptycho.github.io/ptypy/}
    \item py4DSTEM \url{https://py4dstem.readthedocs.io/en/latest/} \cite{savitzky2021py4dstem}
    \item libertem \url{https://libertem.github.io/LiberTEM/} \cite{clausen2020libertem}
    \item FPD \url{https://fpdpy.gitlab.io/fpd/} \cite{nord2020fast, paterson2020fast}
    \item Muller Group codes (Cornell) \url{https://github.com/muller-group-cornell/ptychography}
    \item Maiden Group codes (Sheffield) \url{https://github.com/andyMaiden/SheffieldPtycho}
    \item Ptychoshelves \cite{wakonig2020ptychoshelves}, 
    \item PtychoPy \url{https://github.com/kyuepublic/ptychopy}
\end{itemize}

\section*{References}
\bibliography{refs}
\end{document}